\documentclass[%
 aip,
 amsmath,amssymb,
reprint,%
]{revtex4-1}

\usepackage{graphicx}
\usepackage{dcolumn}
\usepackage{bm}

\usepackage[utf8]{inputenc}
\usepackage[T1]{fontenc}
\usepackage{mathptmx}

\usepackage{etoolbox}
\usepackage{xcolor}
\usepackage{soul}
\makeatletter
\def\@email#1#2{%
 \endgroup
 \patchcmd{\titleblock@produce}
  {\frontmatter@RRAPformat}
  {\frontmatter@RRAPformat{\produce@RRAP{*#1\href{mailto:#2}{#2}}}\frontmatter@RRAPformat}
  {}{}
}%
\makeatother

\begin{document}
\preprint{AIP/123-QED}

\title[3D cylindrical BGK model of electron phase-space holes with finite velocity and polarization drift]{3D cylindrical BGK model of electron phase-space holes with finite velocity and polarization drift}
\author{Ga\"etan Gauthier}
 \email{gaetan.gauthier@lpp.polytechnique.fr}
 \affiliation{Laboratoire de Physique des Plasmas (LPP), CNRS, Sorbonne Universit\'e,
Observatoire de Paris, Universit\'e Paris-Saclay,
Ecole polytechnique, Institut Polytechnique de Paris, 91120 Palaiseau, France}%

\author{Thomas Chust}%
\affiliation{Laboratoire de Physique des Plasmas (LPP), CNRS, Sorbonne Universit\'e,
Observatoire de Paris, Universit\'e Paris-Saclay,
Ecole polytechnique, Institut Polytechnique de Paris, 91120 Palaiseau, France}%

\author{Olivier Le Contel}%
\affiliation{Laboratoire de Physique des Plasmas (LPP), CNRS, Sorbonne Universit\'e,
Observatoire de Paris, Universit\'e Paris-Saclay,
Ecole polytechnique, Institut Polytechnique de Paris, 91120 Palaiseau, France}%

\author{Philippe Savoini}%
\affiliation{Laboratoire de Physique des Plasmas (LPP), CNRS, Sorbonne Universit\'e,
Observatoire de Paris, Universit\'e Paris-Saclay,
Ecole polytechnique, Institut Polytechnique de Paris, 91120 Palaiseau, France}%

\date{\today}

\begin{abstract}
Nonlinear electron kinetic structures are regularly observed in space and experimental magnetized plasmas, called electron phase-space holes (EHs). The existence of EHs is conditioned and varies according to the ambient magnetic field and the parameters of the electron beam(s) that may generate them. The objective of this paper is to extend the 3D Bernstein-Greene-Kruskal (BGK) model with cylindrical geometry developed by Chen et al.\cite{chen2004bernstein, chen2005width} to include simultaneously finite effects due to (i) the strength of the ambient magnetic field $\mathbf{B}_0$, by modifying the Poisson equation with a term derived from the electron polarization current, and (ii) the drift velocity $u_e$ of the background plasma electrons with respect to the EH, by considering velocity-shifted Maxwellian distributions for the boundary conditions. This allows us to more realistically determine the distributions of trapped and passing particles forming the EHs, as well as the width-amplitude relationships for their existence.
\end{abstract}

\maketitle

\section{\label{sec:level1}Introduction}
Electron phase-space holes (EHs) are  non-linear kinetic structures observed in the Earth's magnetosphere by spacecraft missions 
in various regions (\textit{e.g,} the auroral region, the bow-shock, the dayside magnetopause and the geomagnetic tail),\cite{matsumoto1994electrostatic, ergun1998fast, bale1998bipolar, franz2005properties, cattell2005cluster, norgren2015slow, fu2020first, le2017lower, tong2018simultaneous, holmes2018electron, steinvall2019observations,andersson2009new, tao2011model,shustov2021dynamics} as well as in the Saturn's magnetosphere by Cassini mission,\cite{williams2006electrostatic,pickett2015electrostatic}  and very recently in the induced Venus magnetosphere by Parker Solar Probe and Solar Orbiter missions.\cite{malaspina2020plasma,haddid2021Solar}
Moreover, this type of structures can be generated by laboratory experiments such as intense laser,\cite{montgomery2001observation} magnetic reconnection facilities\cite{fox2008laboratory} and by beam injection.\cite{lefebvre2010laboratory} Some theories have been developed for solitary waves in quantum plasma \cite{mamun2010solitary} based on a fluid model or a Bernstein-Greene-Kruskal (hereafter, BGK) approach\cite{bernstein1957exact} for quantum Vlasov equation.\cite{haas2020bernstein} More generally, coherent structures appear in many fields of physics such as biophysics, condensed matter or fluid dynamics.\cite{dauxois2006physics}

First observations in magnetized space plasmas have evidenced electrostatic structures propagating along the ambient magnetic field $\mathbf{B}_0$,\cite{matsumoto1994electrostatic,williams2006electrostatic,krasovsky1997bernstein} although, for fast moving structures, a perpendicular magnetic perturbation  to $\mathbf{B}_0$ ($\delta B_{\perp}$)  corresponding to the Lorentz field was detectable.\cite{ergun1998fast}
Recently, observations in the tail of the Earth's magnetosphere have also reported structures with another electromagnetic signature, characterized by a magnetic field perturbation parallel to $\mathbf{B}_0$ ($\delta B_{\parallel}$)\cite{andersson2009new,tao2011model,holmes2018electron,steinvall2019observations}. These structures associated with electron trapping and a hole in phase space, are characterized in real space by a positive electric potential and an electron density depletion in their center. They therefore have a divergent electric field and their crossing in the direction parallel to $\mathbf{B}_0$ shows a bipolar (monopolar) spike of the parallel (perpendicular) electric field component. Their parallel width is generally in the order of a few to ten Debye lengths, while their perpendicular width can be much larger. The determination of these lengthscales  and their ratio (shape of the structure) are most probably related to the generation mechanism, and to some existence criteria to which this article tries to bring new elements of understanding. EHs can typically be generated during the nonlinear phase of beam instabilities caused by the interaction of several plasma populations.\cite{omura1996electron,mottez2001instabilities,umeda2004two,lu2008perpendicular} In a magnetized plasma, they move along the field line at a velocity $v_{\mathrm{EH}}$ close to the average velocity $v_b$ of the electron beam(s), and are commonly characterized into two types: slow\cite{norgren2015slow} or fast \cite{ergun1998fast}, depending on whether $|u_e| \ll v_{T}$ or $|u_e| \ge v_{T}$, with $u_e$ the drift velocity of the background plasma electrons with respect to the EH and $v_{T}$ the parallel thermal velocity of electrons. As demonstrated numerically by Muschietti et al.\cite{muschietti2000transverse,muschietti2002modeling} and experimentally by Fox et al.,\cite{fox2008laboratory} one of the criteria for the existence of such a nonlinear structure is that the electrons remain sufficiently magnetized to be stable.

In the examples cited above, the plasma conditions vary and, in particular, the magnitude of the ambient magnetic field. Historically, BGK models were developed in 1D\cite{bernstein1957exact,chen2001trapped} for unmagnetized plasma and then theoretical additions  or suggestions were made: shifted distribution \cite{turikov1984electron, chen2005width}, non-maxwellian distribution (\textit{e.g,} flap-top,\cite{muschietti1999phase} Lorentzian\cite{goldman2007theory, chen2005width}), effects of electron polarization current,\cite{franz2000perpendicular, chen2004bernstein} and 2D\cite{ng2006weakly} or 3D\cite{chen2004bernstein, chen2005width} extension. However, these previous studies did not investigate dimensionality and finite velocity effects at the same time,\cite{bernstein1957exact, schamel1979theory, muschietti1999phase, turikov1984electron, chen2005width} and did not give a quantified estimate of electron polarization effects.\cite{franz2000perpendicular, chen2004bernstein} In this paper, we develop a BGK theoretical model that includes the effects of dimensionality, finite ambient magnetic field, and a drift velocity of the background plasma electrons with respect to the EH (\textit{i.e,} 3D axisymmetric cylindrical coordinates, shifted electron distribution function and electron polarization current). It aims at discussing the previous studies,  modeling slow and fast EH observations for relatively weak and strong magnetized plasma, and relating the model to the PIC simulations.\cite{matsumoto1994electrostatic, omura1996electron, umeda2004two, lu2008perpendicular} In particular, the questions addressed in this paper are: how do the ambient magnetic field strength and the finite electron drift velocity influence the existence conditions and characteristics of such 3D EHs? 

After a presentation of the 3D BGK model of EHs with cylindrical geometry, and its underlying assumptions, such as those that allow the inclusion of electron polarization effects (section \ref{sec::BGK}), the characteristics (distribution functions and densities) of the passing and trapped particles are theoretically determined in section \ref{sec::Particles}. Section \ref{sec::Existence} aims at deriving the conditions of existence of these EHs. Our model is then discussed in section \ref{sec::Discussion} in the light of space and laboratory measurements, as well as simulation results.

\section{3D BGK model} \label{sec::BGK}

We consider the case of an uniform magnetized plasma with the assumption of an unperturbed neutralizing ion background. The dynamics of ions is thus ignored for reasons of simplicity and because of their large mass ratio with electrons. The parallel dynamics of electrons (of charge $-e$ and mass $m$) is analyzed from the evolution of their distribution function $f_e$, while  their perpendicular dynamics is described by a fluid representation of their polarization drift.

\subsection{Parallel Vlasov dynamics}

The original BGK model \cite{bernstein1957exact} of EH is a one dimensional,  stationary ($\partial_{t}f_{e}=0$) nonlinear exact solution of the Vlasov-Poisson system of equations for a given shape of potential. Based on the observations \cite{andersson2009new, tao2011model,holmes2018electron, steinvall2019observations, tong2018simultaneous}, the EH potential can be represented in cylindrical coordinates ($r,\theta, z$) independent by rotation around the ambient magnetic field axis ($\mathbf{B}_0 = B_0\hat{\mathbf{z}}$), and can be written in the EH frame in the following double-Gaussian form:
\begin{equation}
    \phi(r,z) = \phi_0\,\exp\left(-\frac{r^2}{2\ell_{\perp}^2} - \frac{z^2}{2\ell_{\parallel}^2}\right) \label{eq::2}
\end{equation}
with $\phi_0$ the amplitude of the potential structure, $\ell_{\perp}$ and $\ell_{\parallel}$ its half-width in the perpendicular and parallel directions to $\mathbf{B}_0$, respectively.
It is also reasonable to consider  the magnetic field perturbation to be much smaller than the ambient magnetic field ($\delta B\ll B_0$) and the perturbed Lorentz term to be negligible compared to the electric field perturbation ($\mathbf{v}\times \delta \mathbf{B} \ll \mathbf{\delta E}$, with $\mathbf{\delta E}=-\mathbf{\nabla}\phi$).
The Larmor radius of the electrons being generally much smaller than the perpendicular size of the EH ($\rho_{L}\ll \ell_{\perp}$) and their travel time in the structure sufficiently slow compared to their gyroperiod ($(|u_e|+v_T)/\ell_{\parallel} \ll \omega_c $),\cite{muschietti2002modeling, franz2000perpendicular, tao2011model, fox2008laboratory} their motion can be approximated by their guiding center motion. Due to the cylindrical symmetry, azimuthal drift displacements ($\mathbf{\delta  E}\times \mathbf{B}_0$) do not contribute to equilibrium in phase space and will not be considered \cite{tao2011model, chen2002GRL}. Indeed, as a first approximation for describing the kinetic dynamics along the ambient magnetic field, we will neglect the radial and azimuthal motions of the electrons (\textit{i.e,} assume no dependence on perpendicular velocities), and we will solve the corresponding simplified Vlasov equation, just along the $z$-axis for a given $r$ value, the electrons being closely tied to a cylindrical magnetic field surface of radius $r$\cite{chen2002GRL, muschietti2002modeling}:
\begin{equation}
    v_z\cdot\frac{\partial f_{e}(r,z,v_z)}{\partial z}+\frac{e}{m}\frac{\partial \phi(r,z)}{\partial z}\cdot\frac{\partial f_{e}(r,z,v_z)}{\partial v_z}=0 \label{eq::1}
\end{equation}
This equation represents a simplified Vlasov equation, in the context of azimuthal symmetry and uniform magnetic field, where the dependence of $\theta$ and perpendicular velocities are neglected. Its conditions of validity deserve some additional comments, which we now address.

The condition to neglect the finite Larmor radius effects on the electrons can be rewritten using the ratio between the electron cyclotron frequency and the electron plasma frequency, and considering the anisotropy ratio of their temperature:
\begin{equation}
    \dfrac{\rho_{L}}{\ell_{\perp}}\ll 1 \quad \Leftrightarrow\quad
     \dfrac{\omega_{p}}{\omega_{c}}\ll \dfrac{\ell_{\perp}}{\lambda_{D}} \sqrt{\dfrac{T_{e\parallel}}{T_{e\perp}}}
     \label{eq::cond1}
\end{equation}
with $\lambda_{D}$ the (parallel) Debye length. This implies for $\omega_{c}/\omega_{p} \leq 1$ that $\ell_{\perp} \gg \lambda_{D}$ (unless $T_{e\perp}\ll T_{e\parallel}$, as in the FAST observation case\cite{ergun1998fast}). The present model has therefore a limit and cannot correctly describe all the structures of small perpendicular size when the plasma is weakly magnetized. In particular, as stressed by Hutchinson\cite{hutchinson2021oblate}, such a model is not valid for EHs with a scale of the order of one Debye length ($\ell _{\perp}\sim \lambda_ D$) in the case of $\omega_p /\omega_c =\rho_{L} /\lambda _D \gg 1$. However, even in the $\ell_{\perp}\sim \rho_{L}$ limit, the guiding center approximation should still give some qualitatively acceptable results.

The finite frequency effects on the electrons are negligible when, in the reference frame moving at their parallel velocity, the time scale of the variations they undergo is much larger than their gyroperiod. 
Similarly, this second condition for the electrons to remain magnetized can be rewritten to give a second condition on the electron cyclotron to plasma frequency ratio:
\begin{equation}
    \dfrac{\omega_p}{\omega_c}\ll \dfrac{ \ell_{\parallel}/\lambda_{D}}{1+|u_e|/v_T} \label{eq::cond2}
\end{equation}
Therefore, for weak magnetic field conditions, small parallel and/or fast velocity structures could hardly be described by our model. 
Note that the previous reasoning also applies to trapped electrons\cite{muschietti1999phase,chen2001trapped} and, insofar as $e\phi_0\leq T_{\parallel}$, implies that their bounce frequency has to be much lower than their cyclotron frequency for the guiding center approximation to be valid:\cite{muschietti2000transverse, muschietti2002modeling}
\begin{equation}
    \omega_b \simeq \dfrac{1}{\ell_{\parallel}} \sqrt{\frac{e\phi_{0}}{m}} \leq \dfrac{v_T}{\ell_{\parallel}} \ll \omega_c \label{eq::cond2trapped}
\end{equation}
Note that this condition is similar to that etablished by Chen et al.,\cite{chen2004bernstein} as given by their Eq. (10). 
However, while we agree that this is a condition for preventing trapped electrons from escaping, i.e. their demagnetization during their bounce motion, we disagree that this condition implies that the effects of polarization drift are negligible. Indeed these latter can be present without causing the running away of electrons.

In view of these limitations, there are however space observations and laboratory measurements for which the conditions developed in this paper are valid and which will be discussed in section \ref{sec::Discussion} (see Table \ref{tab:table1} and references therein).

\subsection{Polarization effects}
\begin{figure*}
    \centering
    \includegraphics[width=\linewidth]{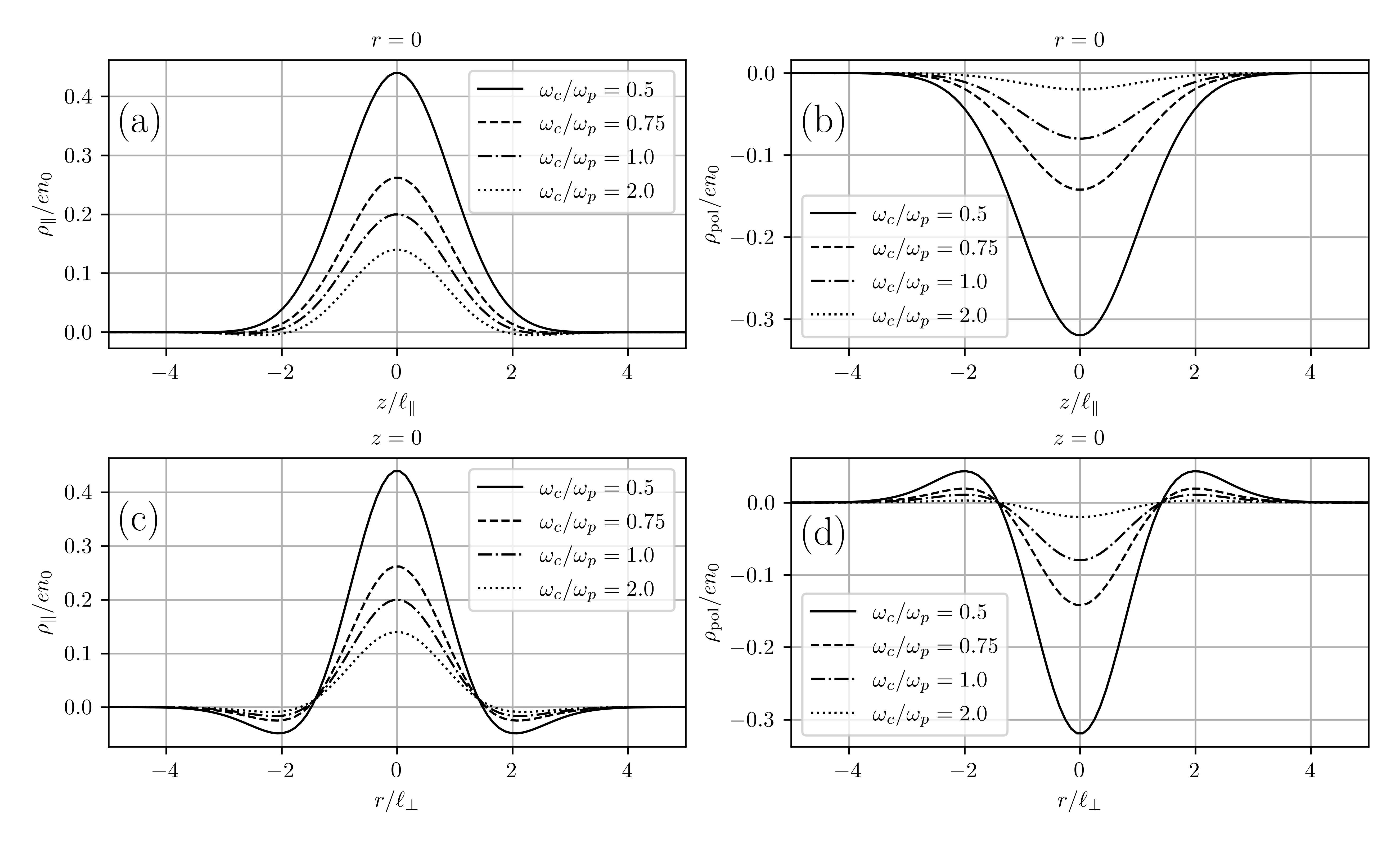}
    \caption{Charge densities $\rho_{\parallel}$ and $\rho_{\mathrm{pol}}$ for different values of cyclotron to plasma frequency ratio $\omega_c/\omega_p$: ($\mathrm{a}$) -- ($\mathrm{b}$) along parallel direction (at $r=0$) and, ($\mathrm{c}$) -- ($\mathrm{d}$) along perpendicular direction (at $z=0$) (with $e\phi_0=T_{e\parallel}$ and $\ell_{\parallel} = \ell_{\perp} = 5\lambda_D$).}
    \label{fig::0}
\end{figure*}

We decompose the electron charge density as a sum of two terms: $\rho_e = \rho_{e\parallel}+ \rho_{e\perp}$, where $\rho_{e\parallel}$ is the contribution to the electron charge density due to their parallel motion, which will be determined from Eq. (\ref{eq::1}); and where $\rho_{e\perp}$ is an additional contribution due to their perpendicular motion. As a direct consequence of azimuthal symmetry, the only perpendicular motion to be considered is the polarization drift. We take into account this effect in the form of an equivalent charge, $\rho_{\mathrm{pol}}=\rho_{e\perp}$, and an additional current in the Maxwell equations. In the reference frame moving at their mean parallel velocity, this electron current is written to lowest order as $\mathbf{J}_{\mathrm{pol}} = n_0m\, \partial_t \mathbf{E}_{\perp}/B_0^2$, where $n_0$ is the electron background plasma density. Taking the divergence:
\begin{equation}
    \mathbf{\nabla}\cdot \mathbf{J}_{\mathrm{pol}}= \frac{\partial }{\partial t}\left[ \mathbf{\nabla} \cdot \left(-\varepsilon_0\frac{\omega_{p}^2}{\omega_{c}^2} \mathbf{\nabla}_{\perp}\phi\right)\right]= - \frac{\partial \rho_{\mathrm{pol}}}{\partial t}
\end{equation}
leads to an expression for the contribution of the electron polarization displacement in the density balance that depends on time. Integrating then with respect to time, this gives us an expression for the electron polarization charge density valid in the EH frame. Hence, the Poisson equation for mobile electrons and infinitely heavy ions,
\begin{equation}
    \varepsilon_0\mathbf{\nabla}^2\phi = -\rho_{\parallel} -\rho_{\mathrm{pol}}
\end{equation}
where $\rho_{\parallel} = en_0 + \rho_{e\parallel}$, can take the following form:\cite{franz2000perpendicular,vasko2017electron,hutchinson2021oblate} 
\begin{equation}
    \mathbf{\nabla}^{2}\phi+\frac{\omega_p^{2}}{\omega_c^{2}}\mathbf{\nabla}_{\perp}^{2}\phi=-\frac{\rho_{\parallel}(r,z)}{\varepsilon_{0}} \label{eq::3}
\end{equation}
The second term on the left-hand side includes thus the perpendicular displacement of electrons due to their finite polarization drift. As Hutchinson\cite{hutchinson2021oblate} pointed out, this term is in principle negligible in the setting where the guiding center approximation applies with structures at Debye lengthscales (as shown in Eqs. (\ref{eq::cond1})-(\ref{eq::cond2}) for $\ell_{\perp}, \ell_{\parallel} \sim \lambda _D$), and should not be invoked for explaining the $\ell_{\perp}/\ell_{\parallel} \simeq (1 + \rho_L^2/\lambda_D^2)^{1/2}$ scaling observed by Franz.\cite{franz2000perpendicular} Nevertheless, with the limitations highlighted in the previous section on the validity conditions of our model, for sufficiently large EHs (a few to a few tens of Debye lengths as estimated from observations), the electron polarization current can play a role in the charge balance, which is evaluated in the next sections.

Using the form of the potential as given by Eq. (\ref{eq::2}), we obtain the following result for the charge densities:
 \begin{equation}
     \frac{\rho_{\mathrm{pol}}}{\varepsilon_{0}}  = \frac{\Lambda-1}{\ell_{\perp}^{2}}\left(\frac{r^{2}}{\ell_{\perp}^{2}}-2\right) \phi \label{eq::4bis}
 \end{equation}
\begin{equation}
     \frac{\rho_{\parallel}}{\varepsilon_{0}}  = \left\{ \frac{1}{\ell_{\parallel}^{2}}\left[\frac{r^{2}}{\ell_{\perp}^{2}}+2\ln\left(\frac{\phi}{\phi_{0}}\right)+1\right]  - \frac{\Lambda}{\ell_{\perp}^{2}}\left(\frac{r^{2}}{\ell_{\perp}^{2}}-2\right) \right\} \phi \label{eq::4}
 \end{equation}
 where $\Lambda = 1+\omega_p^{2}/\omega_c^{2}$. Figure \ref{fig::0} represents the charge densities $\rho_{\parallel}$ and $\rho_{\mathrm{pol}}$ along $z$ and $r$ directions as calculated by Eqs. (\ref{eq::4bis}) -- (\ref{eq::4}). 
 We then observe that the contribution of the polarization current $\rho_{\mathrm{pol}}$ (see Figures \ref{fig::0}(b) and \ref{fig::0}(d)) to the total charge density $\rho_{\mathrm{tot}}=-\varepsilon_0\mathbf{\nabla}^2\phi$ can be of the same order when $\omega_c/\omega_p\leq 1$, which consequently changes significantly the values of $\rho_{\parallel}$ (see Figures \ref{fig::0}(a) and \ref{fig::0}(c)). For consistency, one can easily verify that the total charge of the electron hole is zero.

\subsection{Integral equation}
We introduce the total energy $\epsilon(r,z,v_z)=mv_{z}^2/2-e\phi$, which is a constant of the electron motion along a cylindrical magnetic field surface parameterized by $r$.  We consider two types of electrons with respect to the potential $\phi$: trapped electrons if their total energy $\epsilon$ do not exceed the potential barrier, \textit{i.e,} if $-e\phi_{0}\leq\epsilon\leq0$, and passing electrons such as $\epsilon>0$. This allows us to write the total electron distribution function by introducing two distribution functions $f_t$ and $f_p$ for trapped and passing electrons, respectively, such as:
\begin{equation}
    f_{e}(r,\epsilon)=
    \left\{
  \begin{array}{ c l }
f_{p}(r,\epsilon) & \mathrm{if}\quad\epsilon>0\\
f_{t}(r,\epsilon) &\mathrm{if}\quad-e\phi_{0}\leq\epsilon\leq0
  \end{array} \right.
\end{equation}
Far away from the potential influence (\textit{i.e,} $z\to \pm \infty$), the passing electron distribution must match the boundary conditions: $f_p(r, \epsilon) = f_{\infty}(r, \epsilon)$.  We can then decompose the electron density using $f_t$ and $f_p$ as in the BGK approach:\cite{bernstein1957exact, muschietti1999phase}
\begin{equation}
    \int_{- {V}}^{0}\frac{f_{t}(r,\epsilon)\,\mathrm{d}\epsilon}{\sqrt{2m(\epsilon+ {V})}}= \mathfrak{g}(r, {V})\label{eq::6}
\end{equation}
with $V= e\phi<e\phi_0$ the potential energy and,
\begin{equation}
\mathfrak{g} =  -\frac{\rho_{e\parallel}}{e}- \int_{0}^{+\infty}     \frac{f_{p}(r, { {\epsilon}})\,\mathrm{d} { {\epsilon}}}{\sqrt{2m( { {\epsilon}}+ {V})}} 
 \label{eq::7}
\end{equation}
The set of Eqs. (\ref{eq::6})--(\ref{eq::7}) has the form of an Abel integral equation that can be solved using Laplace techniques \cite{bernstein1957exact,muschietti1999phase} or another method from \S 12 Landau \& Lifshitz,\cite{landau1976mechanics}and considering physical distributions, \textit{i.e,} $\mathfrak{g}(r,0)=0$ : 
\begin{eqnarray}
f_{t}(r,\epsilon) &=& \frac{\sqrt{2m}}{\pi}\int_{0}^{-\epsilon}\frac{\mathrm{d}\mathfrak{g}(r, {V})}{\mathrm{d} {V}}\frac{\mathrm{d} {V}}{\sqrt{-\epsilon- {V}}}\\
&=& f_{t}^{(\mathrm{a})}+ f_{t}^{(\mathrm{b})}  
    \label{eq::7-bis}
\end{eqnarray}
where $f_{t}^{(\mathrm{a})}$ and $f_{t}^{(\mathrm{b})}$ are the integrals corresponding to the  two terms of $\mathfrak{g}$ as given by Eq. (\ref{eq::7}). 
The term $f_{t}^{(\mathrm{a})}$ stands for the electron charge density $\rho_{e\parallel}$,  and $f_{t}^{(\mathrm{b})}$ for the contribution of the passing electrons. 

\section{Signature of trapped and passing electrons} \label{sec::Particles}

\begin{figure*}
    \centering
     \includegraphics[width=\linewidth]{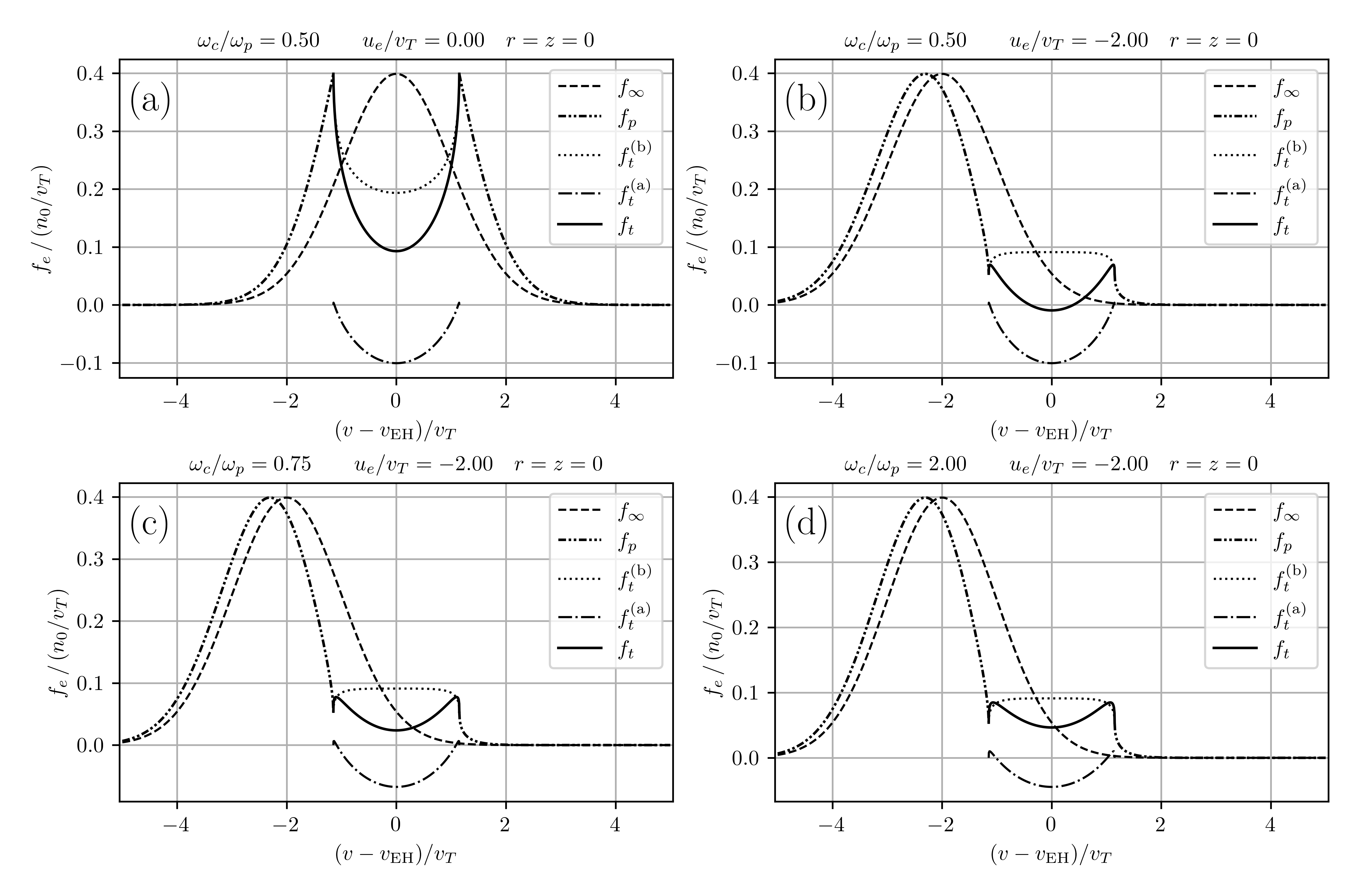}
    \caption{Passing ($f_p$), trapped ($f_t$) electron distribution functions at $r=z=0$, and background electron distribution function ($f_{\infty}$) at infinity, in EH rest frame, for different values of drift velocity $u_e$ and ratio $\omega_c/\omega_p$ (with $e\phi_0 = 0.66 T_{e\parallel}$, $\ell_{\parallel} = 5\lambda_D$, and $\ell_{\perp} = 7\lambda_D$).} 
    \label{fig::1}
\end{figure*} 

\subsection{Distribution functions}
As a first step, using the form of the density as determined by Eq. (\ref{eq::4}), the first integral becomes:
\begin{eqnarray}
    f_{t}^{(\mathrm{a})}(r,\epsilon) & = & \frac{n_0}{v_{T}}\frac{2\sqrt{2}}{\pi} \sqrt{\frac{-\epsilon}{T_{e\parallel}}} 
    \left[-\frac{2\Lambda}{\ell_{\perp}^{2}}+\frac{r^{2}}{\ell_{\perp}^{2}} \left(\frac{\Lambda}{\ell_{\perp}^{2}}-\frac{1}{\ell_{\parallel}^{2}}\right)\right.\nonumber \\  
    && \left.  \quad\quad\quad + \frac{1}{\ell_{\parallel}^{2}} \left( 1-2\ln \left(\frac{-4\epsilon}{e\phi_{0}}\right) \right) \right] \lambda_D^2   
\end{eqnarray}
where $v_{T}=(T_{e\parallel}/m)^{1/2}$ is the parallel thermal velocity of electrons far from the EH, and $\lambda_D =v_{T}/\omega_p$. This term is negative at the bottom of the well, \textit{i.e,} when $\epsilon\to-e\phi_{0}$, then becomes positive and cancels for $\epsilon\to0$. For the case of a negligible electron polarization current, i.e. for $\Lambda=1$, we obtain the same result as Chen et al.\cite{chen2004bernstein}

The passing electrons for $|z|\gg \ell_{\parallel}$ are all corresponding to the background plasma. Thus, if at infinity their distribution $f_{\infty}$ is a velocity shifted Maxwellian distribution (independent of $r$), everywhere else it must be of the following form:  
\begin{equation}
    f_{p}(\epsilon) = \frac{n_{0}}{\sqrt{2\pi} \, v_{T}} \sum_{\sigma=\pm1} \exp \left[-\frac{(\sigma\sqrt{2\epsilon/m}-u_{e})^{2}}{2v_{T}^{2}}\right] 
\end{equation}
with $\epsilon>0$ and $u_{e}$ the drift velocity of the background plasma with respect to the EH. The contribution of passing electrons to the distribution of trapped electrons can then be written as: 
\begin{eqnarray}
    f_{t}^{(\mathrm{b})}(\epsilon) 
    &=&\frac{2\,n_{0}}{\pi\sqrt{2\pi} \, v_{T}}\left[{I}(\beta,\zeta)+ {I}(-\beta,\zeta)\right]  \label{eq::13a}
\end{eqnarray}
with
\begin{eqnarray}
    {I}(a, b)=\int_{0}^{+\infty}\frac{\mathrm{e}^{-(a x - b)^{2}}}{1+x^{2}}\mathrm{d}x \label{eq::13}
\end{eqnarray}
and where we have defined the quantities $\beta = (-\epsilon/T_{e\parallel})^{1/2}$ and $\zeta=u_{e}/\sqrt{2}v_{T}$. As an integral function of a positive integrand, we get a positive function. The general expression (\ref{eq::13}) does not allow an analytical calculation, and must be evaluated numerically, except when the integral takes the form of the Dawson's integral, for $b=0$. In that case of zero drift velocity ($u_e=0$), corresponding to the EH immobile with respect to the background plasma, we get:
\begin{equation}
    f_{t}^{(\mathrm{b})} (\epsilon) = \frac{n_0}{v_T}\,\sqrt{\frac{2}{\pi}} \exp(-\beta)\left[1-\mathrm{erf }(\sqrt{-\beta})\right]
\end{equation}
where the $\mathrm{erf}$ function is defined as $\sqrt{\pi}\,\mathrm{erf}(x)/2= \int_0^x\exp(-t^2)\mathrm{d}t$. For the case of zero drift velocity, we then well find the result obtained by previous authors, whether in the 1D\cite{turikov1984electron, chen2001trapped} or 3D\cite{chen2002GRL, chen2004bernstein} BGK model. For the case of a finite drift velocity, the solution (\ref{eq::13a}) is in accordance with Turikov's result.\cite{turikov1984electron} The continuity property of this general solution when $\epsilon\to0$ can be easily verified. First, we obtain straightforwardly $f_t^{\mathrm{(b)}}(r,\epsilon\to0^-)=f_{p} (r,\epsilon\to0^+)$. Secondly, since $f_t^{\mathrm{(a)}}(r,\epsilon\to 0^-)=0$ we get the continuity relation $f_e(r,\epsilon\to0^-)=f_e(r,\epsilon\to0^+)$, which is consistent with the populations present outside the hole and generation mechanism.

Figure \ref{fig::1} displays the different contributions to the distribution of trapped electrons ($f_t$, with the virtual terms $f_t^{(\mathrm{a})}$ and $f_t^{(\mathrm{b})}$ determined above), the distribution of passing electrons ($f_p$), and the electron distribution at infinity ($f_{\infty}$), as a function of velocity in the EH reference frame. The drift-free case ($u_e=0$) is represented by the Figure \ref{fig::1}(a) as reference. The impact of $u_e$ is visualized by comparing to the Figure \ref{fig::1}(b), which represents a case with finite drift ($u_e = -2v_T$). We observe an asymmetric distribution, an important decrease in the positive virtual value of $f_t^{\mathrm{(b)}}$ without any change in the equally virtual term associated with the potential, $f_t^{\mathrm{(a)}}$. As a result, in that case with $\omega_c/\omega_p = 0.5$, the distribution of trapped electrons $f_t$ is found to be slightly negative, which is unphysical. As we shall see later, the conditions for the existence of EHs will be defined on the basis of this limitation. In addition, the pseudo-Maxwellian part of $f_p$ is moved at higher velocities, allowing for particle acceleration. Comparing Figures \ref{fig::1}(b)--(d) shows the impact of the drift polarization effect on $f_t^{\mathrm{(a)}}$ (which increases with $\Lambda$, \textit{i.e,} decreases with $\omega_c/\omega_p$),  without modifying $f_t^{\mathrm{(b)}}$. As $\Lambda$ increases, there is a decrease in the density of trapped electrons $f_t$ at the hole center ($v=v_{\mathrm{EH}}$). 

Figure \ref{fig::1}(a) with $u_e \ll v_T$ represents the case of a slow EH  (\textit{e.g,} as observed in the magnetotail\cite{norgren2015slow, fu2020first}) that may result from  counterstreaming instability.\cite{goldman2000turbulence, mottez2001instabilities, umeda2006nonlinear} Conversely,  Figures \ref{fig::1}(c)--(d) with $u_e \geq v_T$ represent the case of fast EH and show a plateau-like structure as observed in space plasmas,\cite{holmes2018electron} or in Particle-In-Cell (PIC) simulations of bump-on-tail type.\cite{omura1996electron, 
umeda2004two, lu2008perpendicular}

The choice of a drift velocity $u_e$ as high as $2 v_T$ may raise questions for a reader accustomed to Schamel's results, showing that values greater than about 1.3 are impossible under physical conditions.\cite{schamel1986tutorial} Unlike the BGK integral equation method we used to solve the Vlasov-Poison equation system, Schamel's approach, which used a differential equation method, leads to a limited number of possible solutions. Indeed, keen to find "preferred" BGK states, Schamel introduced a particular shape of the trapped distribution function, which is smoother at the separatrix $\epsilon = 0$ than most BGK solutions, and in particular ours as can be seen on Figure \ref{fig::1}. Justifying what might be an acceptable trapped distribution function remains an open question\cite{hutchinson2017electron} and is beyond the scope of our work. However, we can acknowledge that such structures have long been observed in space and laboratory plasmas. For instance, Andersson et al.\cite{andersson2009new} reported fast EHs in the Earth’s magnetotail with velocities $u_e \ge 2.5$, whereas histogram given by Holmes et al.\cite{holmes2018electron} show velocities between 1 and 1.5. In addition, Fox et al.\cite{fox2008laboratory} reported EHs with velocities $u_e \sim 2$  under laboratory conditions, as did Lefebvre et al.,\cite{lefebvre2010laboratory} who found 80\% of structures with velocities between 1.3 and 2.3, with 2.1 as the median velocity.

\subsection{Electron densities in the hole
\label{sec::electrondensities}}

From the passing distribution function $f_p$, we can explicitly define the  density $n_p$ of passing electrons as:
\begin{eqnarray} \nonumber
    n_p(r,z)&=& \frac{n_0}{\sqrt{2\pi}\,v_T}\left( \int_{+v_T\sqrt{2\psi}}^{+\infty}
    \exp{\left[-\frac{(\sqrt{v^2-\frac{2e\phi}{m}}-u_e)^2}{2v_T^2}\right]}\mathrm{d}v\right.\\ \nonumber &&+ \left. \int^{-v_T\sqrt{2\psi}}_{-\infty}
    \exp{\left[-\frac{(-\sqrt{v^2-\frac{2e\phi}{m}}-u_e)^2}{2v_T^2}\right]}\mathrm{d}v \right)\\
    &=& n_0\left[{J}_{+}(\sqrt{2\psi}, \zeta)+{J}_{-}(\sqrt{2\psi}, \zeta) \right] \label{eq::pdensity}
\end{eqnarray} 
with
\begin{equation} \label{integralJ}
 {J}_{\sigma}(a, b) = \frac{1}{\sqrt{2\pi}}\int_{a}^{+\infty} \!\!\!\mathrm{e}^{-(\sigma \sqrt{x^2-a^2}- \sqrt{2}b)^2/2}\mathrm{d}x    
\end{equation}        
and where $\psi = e\phi/T_{e\parallel}$, the ratio between the potential and thermal energy of electrons. In particular for zero drift velocity ($u_e =0$), ${J}_{+}(\sqrt{2\psi},0) = {J}_{-}(\sqrt{2\psi},0) = \exp(\psi)\left[1-\mathrm{erf}(\sqrt{\psi})\right]/2$, whence:
\begin{equation}
    n_p(r,z) = n_0 \exp(\psi)\left[1-\mathrm{erf}(\sqrt{\psi})\right]\leq n_0
\end{equation}
For finite drift velocity, the integral (\ref{integralJ}) must be evaluated numerically. Using the definition of the charge density $\rho_{\parallel}$, the trapped electron density $n_t$ writes:
\begin{equation}
    n_t(r,z) = n_0 - n_p(r,z) - \frac{\rho_{\parallel}(r,z)}{e} \label{eq::19}
\end{equation} 
and can be determined from Eqs. (\ref{eq::4}) and (\ref{eq::pdensity}). In the case where $u_e = 0$ and $\Lambda=1$ (negligible effects of EH velocity and electron polarization drift), the trapped density $n_t$ calculated by Chen et al.\cite{chen2004bernstein,chen2005width} is retrieved. We can also define the density of the trapped electrons from their distribution as:  
\begin{equation}
n_t(r,z) = \int_{-v_T\sqrt{2\psi}}^{v_T\sqrt{2\psi}} f_t(r,z,v) \mathrm{d}v
\end{equation}

\begin{figure}
    \centering    
    \includegraphics[width=1.03\linewidth]{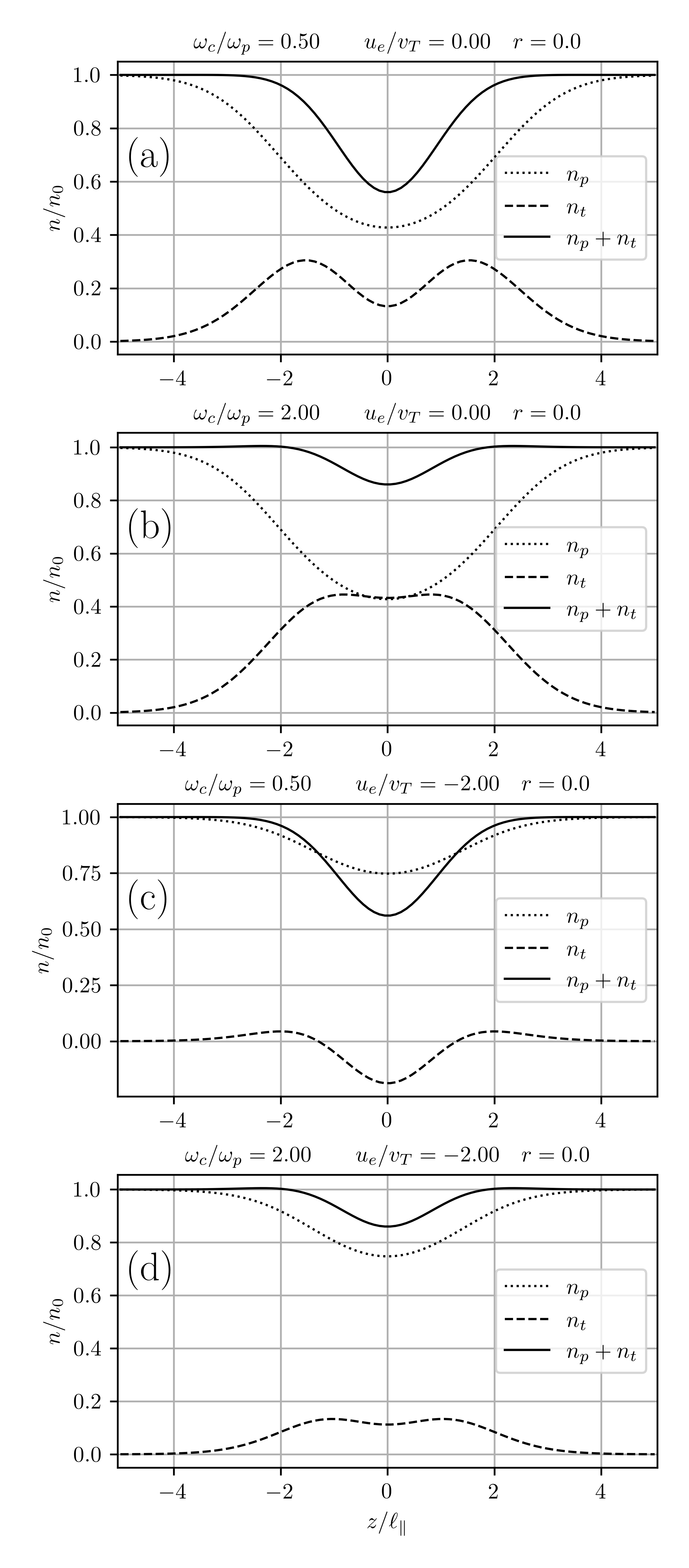}     
    \caption{Passing ($n_p$), trapped ($n_t$) electron densities along parallel direction, at $r=0$, for different values of cyclotron to plasma frequency ratio $\omega_c/\omega_p$ and drift velocity $u_e$ (with $e\phi_0=T_{e\parallel}$ and $\ell_{\parallel} = \ell_{\perp} = 5\lambda_D$).}
    \label{fig::1-bis}
\end{figure}

Figure \ref{fig::1-bis} displays the different electron densities in the hole ($n_t$, $n_p$ and their sum $n_t+n_p$) as a function of parallel axis $z$. As indicated by the Poisson Eq. (\ref{eq::3}) and Eq. (\ref{eq::19}), the quantity $n_p+n_t$ represents the electron density $-\rho_{e\parallel}/e$ induced by the potential of the structure and the perpendicular polarization drift effect. It can be observed that despite the variations of the quantities $n_p$ and $n_t$, the overal neutrality remains preserved and the total load of the structure is zero. The impact of the polarization effect can be seen by comparing Figures \ref{fig::1-bis}(a) and \ref{fig::1-bis}(b). We observe that this leads to a reduction of the density of electrons trapped in the potential well, and more importantly at the center than at the edges (dashed line in Figure \ref{fig::1-bis}(a)). Thus, for $\omega_c/\omega_p < 1$, the distribution of trapped electrons consists of two humps. This is because the polarization current brings additional electrons to the center of the structure (see Figures \ref{fig::0}(b) and \ref{fig::0}(d)), independently of the parallel electron dynamics, so fewer trapped electrons are needed to satisfy Poisson's equation. Figures \ref{fig::1-bis}(b) and \ref{fig::1-bis}(d), or \ref{fig::1-bis}(a) and \ref{fig::1-bis}(c), illustrate the impact of the drift velocity. As it results in more passing electrons, there is also less need for trapped electrons. For $\omega_c/\omega_p=0.5$, $u_e=-2v_T$ and assuming a spherical shape ($\ell_{\parallel}= \ell_{\perp} =5\lambda_D$), as shown in Figure \ref{fig::1-bis}(c), the case is even impossible, as it would require a non-physical negative density of the trapped electrons. In this exemple, theoretical determination of the distribution of trapped electrons ($f_t$) in fact reveal negative values around its center, similar to what is observed in Figure \ref{fig::1}(b). We can note that the lowest order approximation of $\mathbf{J}_{\mathrm{pol}}$ we have used, is in principle strictly consistent for small parallel perturbations of the electron density, \textit{i.e,} $n_t+n_p\sim  n_0$. For this reason, all our evaluations carried out with strong disturbances are to be considered with caution and aim at identifying trends.

This presentation of the electron density composition in the EH allows us to recall an important observation made by Chen and Parks\cite{chen2001trapped, chen2002GRL} on the nature of the charge density ``shielding'' of the core of a BGK structure. As Figure 3 shows, the passing electrons in the middle of the hole are in deficit (relative to the ions), simply because they are accelerated there. The positive core is then ``shielded'' by the trapped electrons that oscillate in the potential structure. In fact, the trapped electrons must distribute themselves in such a way as to counterbalance the positive charge density produced by the depletion of passing electrons inside the potential well, thereby producing a total charge density consistent with the specified potential profile. A BGK EH is a self-consistent and self-sustaining object with zero total charge and does not require any thermal screening by the surrounding plasma (Debye shielding). This result contradicts the idea that the positive core of the EH is due to a deficit of trapped electrons, and that this positive core is screened by the 
passing electrons, as for example recently described by Hutchinson.\cite{hutchinson2021oblate}

\section{Existence criteria \label{sec::Existence}}

\begin{figure*}
    \centering    
    \includegraphics[width=1.02\linewidth]{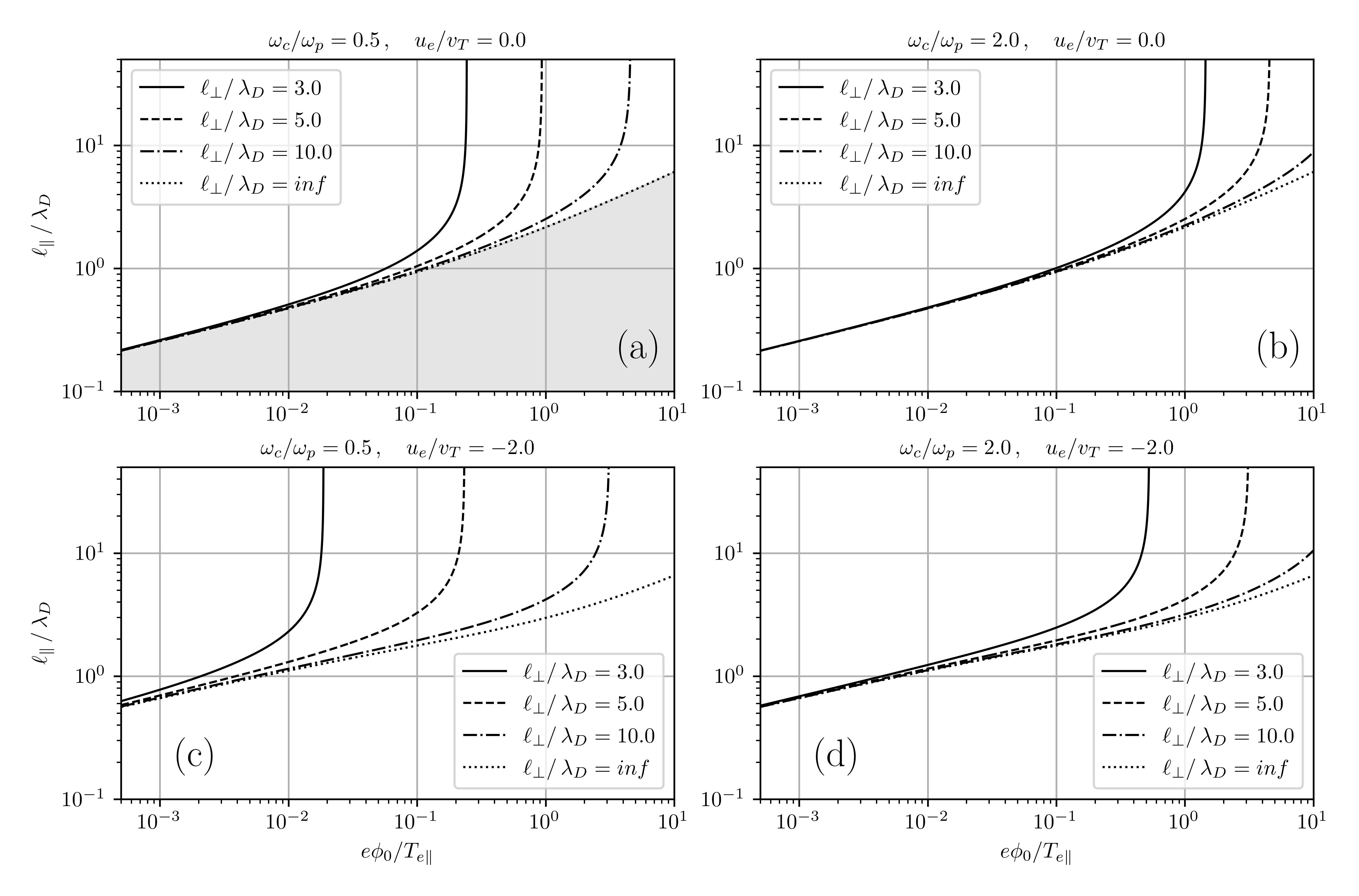}
    \caption{Parallel width-amplitude relations for different values of perpendicular width $\ell_{\perp}$, drift velocity $u_e$ and cyclotron to plasma frequency ratio $\omega_c/\omega_p$.}
    \label{fig::2}
\end{figure*}

 We define and analyse the criteria that allow these EH structures to exist in order to obtain information on their size. The trapped distribution function (\ref{eq::7-bis}) has to be physical, and thus $f_{t}$ must be non-negative. Writing $f_t(\epsilon)\geq 0$ for $-e \phi_0 \leq \epsilon <0$, we obtain: 
\begin{equation}
    \frac{\ell_{\parallel}^2}{\lambda_D^2} \geq \frac{2\ln4-1}{ {G}(\psi_0,\zeta)-2\Lambda \lambda_D^2/\ell_{\perp}^2} \label{eq::8}
\end{equation}
where $\psi_0 = e\phi_0/T_{e\parallel}$ is the ratio $\psi$ at $r=z=0$ corresponding to the potential maximum,
and where we have introduced the function 
\begin{equation}
     {G}(\psi_0,\zeta) = \frac{ {I}(\sqrt{\psi_0},\zeta)+ {I}(-\sqrt{\psi_0},\zeta)}{2\sqrt{\pi \psi_0}}
\end{equation}
 This function has the following limits, for $\psi_0 \to 0^+$: $ {G}(\psi_0,\zeta)\sim\sqrt{\pi/4\psi_0}\exp(-\zeta^2)$, and for $\psi_0 \to +\infty$: $ {G}(\psi_0,\zeta)\sim 1/2\psi_0$. The case without drift velocity ($u_e=0$) gives:
 \begin{equation}
      {G}(\psi_0,0) =\frac{\sqrt{\pi}}{2\sqrt{\psi_0}} \exp(\psi_0)[1-\mathrm{erf}(\sqrt{\psi_0})]
 \end{equation}
The sign of the denominator in Eq. (\ref{eq::8}) imposes a second condition on the perpendicular scale,  which must be respected and can be expressed as:
\begin{equation}
    \frac{\ell_{\perp}^2}{\lambda_D^2}\geq \frac{2\Lambda}{ {G}(\psi_0,\zeta)}
\label{eq::9}
\end{equation}

Eqs. (\ref{eq::8}) and (\ref{eq::9}) give amplitude-width criteria in both parallel  and perpendicular directions. In the case where $u_e=0$ and $\Lambda=1$, we obtain the same criteria as those found by Chen et al.\cite{chen2004bernstein,chen2005width} Figure \ref{fig::2}  represents the minimum parallel sizes of the hole (from Eq. (\ref{eq::8})) as a function of the amplitude of the potential well for several values of the perpendicular size, and plasma parameters.
While the observations indicate that the ratio $e\phi_0/T_{e\parallel}$ does not seem to exceed 1 by much, we have extended its range to 10 in order to show the common trend for the different values of $\ell_{\perp}$. The area under these curves represents forbidden zones, see for example the shaded area of Figure \ref{fig::2}(a) corresponding to the case $\ell_{\perp}=\infty$, \textit{i.e,} the 1D limit.
\begin{figure}
    \centering    
    \includegraphics[width=1.04\linewidth]{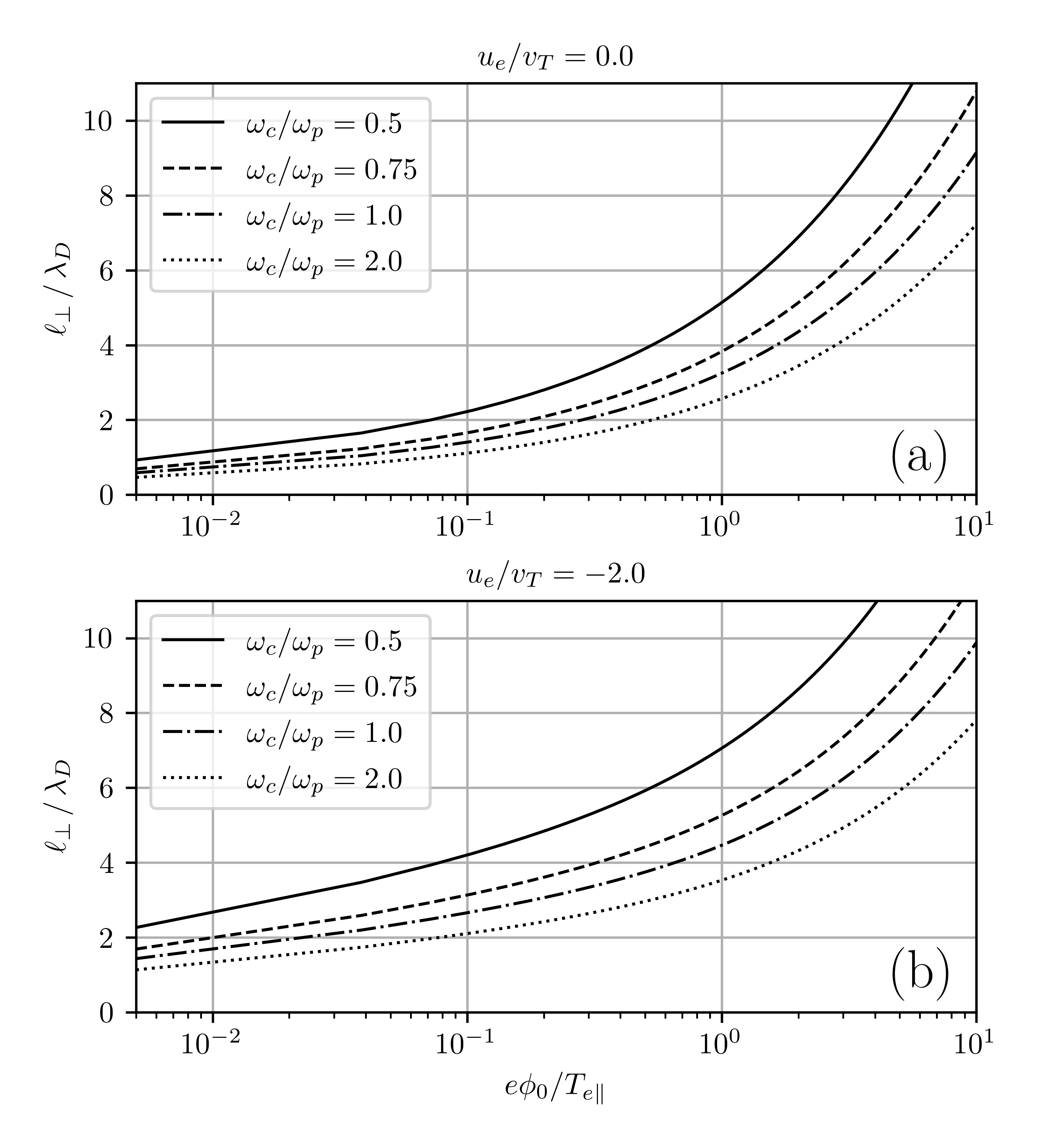}
    \caption{Perpendicular width-amplitude relations for different values of  cyclotron to plasma frequency ratio $\omega_c/\omega_p$ and drift velocity $u_e$.}
    \label{fig::3}
\end{figure}
We note that for a finite perpendicular length the weaker the magnetic field strength or the higher the electron drift velocity, the larger the forbidden zone. Indeed, polarization drift effects tend to reduce the possibility of perpendicular  small scale, large amplitude structures (compare Figures \ref{fig::2}(a) and \ref{fig::2}(b) where $u_e=0$). When a shifted Maxwellian distribution is taken into account, which should be more realistic regarding to a beam instability generation mechanism, we find that the allowed zone is still more reduced, making structures below the Debye lengthscale in the parallel direction hardly possible (compare Figures \ref{fig::2}(b) and \ref{fig::2}(d)). This last result is consistent with Goldmann's.\cite{goldman2007theory} Furthermore, when the finite effects of electron polarization current and electron drift velocity are both considered, perpendicular small scale structures can only exist for very small amplitudes (see Figure \ref{fig::2}(c)). 

 Figure \ref{fig::3} represents the minimum perpendicular sizes of the hole (from Eq. (\ref{eq::9})) as a function of amplitude of the potential well for several  $\omega_c/\omega_p$ ratio, in case of $u_e=0$ (Figure \ref{fig::3}(a)) and $u_e = -2v_T$ (Figure \ref{fig::3}(b)). For both cases, the polarization drift effects lead to an increase in the slope of $\ell_{\perp}(\psi_0)$ and of the forbidden zone for small $\omega_c/\omega_p$ values. For a given $\omega_c/\omega_p$ ratio, the velocity drift effects also contribute significantly to the expansion of the forbidden zone. In addition, the width-amplitude relations $\ell_{\perp}(\psi_0)$ shown in Figure \ref{fig::3} can also represent the maximum well amplitudes $\psi_{0,\max}$ as a function of $\ell_{\perp}$. The vertical asymptotes in Figure \ref{fig::2} correspond to the limit of $\ell_{\parallel}(\psi_0\to \psi_{0,\max})$, precisely. Consistently, from Figure \ref{fig::2}(c), we observe that the conditions used to make Figure \ref{fig::1-bis}(c) (\textit{i.e,} $e\phi_0=T_{e\parallel}$ and $\ell_{\parallel} = \ell_{\perp} = 5\lambda_D$, with $u_e = -2v_T$, and  $\omega_c/\omega_p = 0.5$) are in the forbidden zone.
 
\section{Discussion}
\label{sec::Discussion}
\begin{table*}
\caption{\label{tab:table1} Experimental data in different regions}
\begin{ruledtabular}
\begin{tabular}{llccccc}
\textit{Article} & \textit{Context} & $\omega_c/\omega_p$ & $\ell_{\parallel}/\lambda_D$ & $\ell_{\parallel}/\ell_{\perp}$ & $e\phi_0/T_{e\parallel}$ &  $|u_e|/v_T$\\
\hline Ergun 1998\cite{ergun1998fast} & Auroral region (FAST) &  $5-15$ & $0.5-4$ & < 1 & $0.05-1.1$  & < 0.6 \\ 
Franz 2005\cite{franz2005properties} & PS/PSBL region (POLAR) & $\sim$ 2  & $0.1-6$&  > 1 & $10^{-5}-10^{-2}$ & $0.3-1$\\ 
Franz 2005\cite{franz2005properties} & Cusp region (POLAR) & $<1$  & $0.5-10$&  <1 & $10^{-3}-10^{-1}$ & $0.1-2$\\ 
Andersson 2009\cite{andersson2009new} & PS region (THEMIS)  &  $\sim 0.8$ & $\sim 15$ &$>1$  &  $\sim 0.5$ & $1.2-5$ \\
Norgren 2015\cite{norgren2015slow} &  PSBL region (Cluster)  &  $\sim 0.4$ & $2-4$ &$\le 0.5$ & 0.1 & $\sim 0.03$ \\
Le Contel 2017 \cite{le2017lower} & PS region (MMS) & $\sim 0.8$  & $\sim 10$ & $0.3-1$ & 0.25 & $0.7-1.8$\\
Holmes 2018 \cite{holmes2018electron} &  Duskside ﬂank region (MMS)  &  $\sim 0.6$ & $4-11$&  $0.3-2$ & $0.5-3$& $1-1.5$ \\
Steinvall 2019 \cite{steinvall2019observations} & PS/PSBL region (MMS) & $\sim 0.5$  & $\sim 10$ & $\sim 0.6$ & $1-2$& $0.6-2$  \\
Fu 2020 \cite{fu2020first} & PS region (MMS) & $\sim 0.13$  & $\sim 5$ & $< 0.5$ & $\sim 0.05$ & $\sim 0.05$ \\
Fox 2008\cite{fox2008laboratory} & Experimental Setup &  $\sim 0.14$ & $\sim 25$ &$\sim 0.5$  & $\sim 1$  & $\sim 2$ \\
Lefebvre 2010 \cite{lefebvre2010laboratory} & Experimental Setup &  0.5-7 & $4.5-24.4$ & not measured  & $0.1-0.75$  & $1.3-2.3$ \\
\end{tabular}
\end{ruledtabular}
\end{table*}

Our study allows to precise the 3D EH existence conditions depending on the electron drift velocity ($u_e$) and the plasma magnetization ($\omega _c/\omega _p$). In this section, we compare our results with various measurements from space and laboratory summarized in Table~\ref{tab:table1}. For instance, the first statistical results of EHs observed by the Fast mission \cite{ergun1998fast} in highly magnetized auroral plasma ($\omega_c/\omega_p>5$) and moving with velocities $u_e<0.6 v_T$, have interpreted  the observational relationship between $e\phi_0/T_e$ and $\ell_{\parallel}/\lambda_D$  from a 1D BGK model.\cite{muschietti1999phase} Most of these observations ($e\phi_0/T_e \approx 0.05-1.1$, $\ell_{\parallel}/\lambda_D \approx 0.5-4$) belong to the allowed region for EH having $\ell_{\perp}/\lambda_D \leq 3$ (Figure \ref{fig::2}(b)). 
As the same manner, a statistical study of EHs detected in the cusp and in the Plasma Sheet (PS)/Plasma Sheet Boundary Layer (PSBL) regions by the Polar satellite, showed a relationship between their potential amplitude and parallel size.\cite{franz2005properties} 
In the cusp region, $\omega_c/\omega_p<1$ (resp. PS/PSBL region, $\omega_c/\omega_p\sim 2$), the EH velocity was found in the range 0.1-2 (resp. 0.3-1) of the thermal velocity. The potential $e\phi_0/T_e$ was found between $10^{-3}$ and $10^{-1}$ (resp. between $10^{-5}$ and $10^{-2}$). Considering these different EH velocities and  $\omega_c/\omega_p$ ratios, the cusp (resp. PS/PSBL) observations can be related to our results in Figures \ref{fig::2}(a) and \ref{fig::2}(c) (resp. \ref{fig::2}(b) and \ref{fig::2}(d))  for the slow and fast EHs. Smallest values of $\ell_{\parallel}/\lambda_D \sim 0.5$ (resp. $\ell_{\parallel}/\lambda_D \sim 0.1$) in the cusp (resp. in the PS/PSBL) region are allowed for the smallest values of the potential $10^{-3}$ (resp. $10^{-5}$, not shown) and only for slow EH speed $u_e/v_T=0$. Fastest EHs in the cusp (resp. PS/PSBL) region with $u_e/v_T=2$ (resp. $u_e/v_T=1$) are allowed only for $\ell_{\parallel}/\lambda_D>0.6$.
Super slow EHs with weak potential ($u_e/v_T \sim 0.05$, $e\phi_0/T_e \sim 0.05-0.1$) reported from  Cluster\cite{norgren2015slow} and MMS\cite{fu2020first} observations in weakly magnetized plasma having $\ell_{\parallel}/\lambda_D \sim 2-5$ are allowed if  $\ell_{\perp}\ge 3$, which is consistent with the observed scale ratio ($\ell_{\parallel}/\ell_{\perp}<0.5$), see  Figure \ref{fig::2}(a). 
From  other observations provided by MMS\cite{ holmes2018electron, steinvall2019observations} in weakly magnetized plasma ($\omega_c / \omega_p \sim 0.5$), fast and large-amplitude EHs ($u_e/v_T \ge 1$, $e\phi_0/T_e \sim 1$) have been also reported. In such conditions and in accordance with observations, these structures are allowed if $\ell_{\perp}\geq 10 \lambda_D$ with a minimum allowed value of $\ell_{\parallel} \sim 4 \lambda_D$, see Figure \ref{fig::2}(c). 

Regarding laboratory measurements, our model is also consistent with the EH properties reported so far. For instance, Fox et al.\cite{fox2008laboratory} observed in a weakly magnetized plasma ($\omega_c/\omega_p \sim 0.14$) fast, large amplitude EHs ($u_e /v_T \sim 2$ and $e\phi_0/T_e \sim 1$) with $\ell_{\parallel}/\lambda_D \sim 25$ and $\ell_{\parallel}/\ell_{\perp} \sim 0.5 $, which are allowed by our model for $\ell_{\perp}/\lambda_D \ge 10$,  as shown in Figure \ref{fig::2}(c). Measurements reported by Lefebvre et al. of fast EHs ($u_e/v_T \sim 1.3-2.3$) in both weakly and stronlgy magnetized plasma ($\omega_c/\omega_p \sim 0.5-7$) also support our results. They have moderate amplitudes ($e\phi_0/T_e \sim 0.1-0.75$) with $\ell_{\parallel}/\lambda_D \sim 4.5-24.4$. In the weakly magnetized regime (Figure \ref{fig::2}(c)), these fast EHs, for potential $\sim 0.1$, are allowed if $\ell_{\perp}/\lambda_D \ge 5$,  and for larger potentials between 0.2 and 0.75, if $\ell_{\perp}/\lambda_D \ge 10$. In the strongly magnetized regime (Figure \ref{fig::2}(d), the constraints are a bit looser, the EHs are allowed with possible smaller perpendicular scales  ($\ell_{\perp}/\lambda_D \ge 3$, for potential $\sim 0.1$; $\ell_{\perp}/\lambda_D \ge 5$, for larger potentials between 0.2 and 0.75). These constraints are more precise than those given by Chen et al.\cite{chen2005width} and considered in Lefevbre et al.\cite{lefebvre2010laboratory}.

Now we compare our results with 2D PIC simulations. Note that our model is developed using a cylindrical geometry with related assumptions whereas simulations use cartesian geometries. Therefore, 
differences could be found depending on the geometry used. However, these should be negligible when the radius of curvature of the cylindrical structure ($\ell_{\perp}$) is much larger than the Larmor radius, which is consistent with the guiding center approximation used. Performing counterstreaming simulations ($u_e\simeq 0$) in different magnetization and amplitude conditions ($\omega_c/\omega_p \simeq 0.5-10$, $e\phi_0/T_{e\parallel} \simeq 0.7-0.8$ or $e\phi_0/T_{e\parallel} \simeq 4$), Umeda\cite{umeda2006nonlinear} obtained both 1D and 2D structures. For $\omega_c/\omega_p \simeq 1-10$,  quasi 1D EHs ($\ell_{\perp}/\lambda_D >128$) are found whereas for a less magnetized plasma ($\omega_c/\omega_p=0.5$) and depending on the velocity beams,  EHs have a 2D structure (\textit{e.g,} $\ell_{\perp}/\lambda_D \simeq 6-8$, $\ell_{\parallel}/\lambda_D\simeq 3-4$). From our model (see Figure \ref{fig::2}(b) where $\omega_c/\omega_p>1$), we observe that 1D EHs ($\ell_{\perp}/\lambda_D = inf$) must have  $\ell_{\parallel}/\lambda_D \geq 2$ for $e\phi_0/T_{e\parallel}\simeq 0.7-0.8$ and $\ell_{\parallel}/\lambda_D>4$ for $e\phi_0/T_{e\parallel}\simeq  3-4$. In the case of 2D EHs found, i.e. for finite values of $\ell_{\perp}/\lambda_D$, the perpendicular width-amplitude relations, shown in Figure \ref{fig::3}(a) for less magnetized plasma, require  $\ell_{\perp}/\lambda_D > 4$  for $e\phi_0 /T_{e\parallel} \simeq 0.7-0.8$, and $\ell_{\perp}/\lambda_D > 5-9$ for $e\phi_0/T_{e\parallel} \simeq 1-4$. 
From the parallel width-amplitude relations shown in Figure \ref{fig::2}(a), these structures must also have $\ell_{\parallel}/\lambda_D > 2$ and $\ell_{\parallel}/\lambda_D > 2-4$, respectively. 
Investigating numerically counterstreaming instability ($u_e\sim0$) in the auroral region ($\omega_c/\omega_p = 5$),  Oppenheim\cite{oppenheim1999evolution} found oblate EHs with $l_{\parallel} = 20~\lambda_D$,  $\ell_{\perp} \sim 100~\lambda_D$, and $e\phi_0/T_{e\parallel}\sim 1$. These structures are finally found unstable associated with the growth of electrostatic whistler waves after thousands of plasma periods. Such quasi-1D structures are consistent with our model, which requires only the condition: $\ell_{\parallel}/\lambda_D \geq 2$  (Fig. \ref{fig::2}(b)), yet indicating that all shapes of structures may  exist. Studying the bump-on-tail instability in the magnetotail ($u_e/ v_T>1$, $\omega_c/\omega_p=1$ and $e\phi_0/T_{e\parallel} \simeq 0.5$), Umeda\cite{umeda2004two} found EHs with $\ell_{\parallel}\simeq 20\lambda_D$ and $\ell_{\perp} > \ell_{\parallel}$ (notably due to coalescence of EHs in the nonlinear phase). Furthermore, the authors found that in such conditions EHs are stable for long time ($\omega_pt\sim 1000$). Based on Figures. \ref{fig::2}(d) and \ref{fig::3}(b), our conditions of existence allow even smaller EHs with $\ell_{\parallel}/\lambda_D > 2-3$  and $\ell_{\perp}/\lambda_D>3$, as well as all shapes of structures. Therefore, all EH structures obtained by numerical simulations are located in the regions of existence of our model.

\section{Conclusions}

To conclude, the model presented in this paper describes the criteria of existence of 3D cylindrical EHs including both the polarization drift of electrons ($\omega_c/\omega_p \lesssim 1$) and a finite velocity of the EH with respect to the background electrons ($|u_e| \gtrsim v_T$). For fast EHs, this allowed us to use more realistic boundary conditions on the electron distribution functions, which include a global drift consistent to the observations (\textit{e.g,} MMS recent observations \cite{holmes2018electron,steinvall2019observations}). These theoretical improvements have shown that they could have an important impact on the  distribution functions and densities of electrons passing through and trapped in the EH structures, as well as on their conditions of existence. These two effects, the polarization drift of electrons and a finite parallel shift in their velocity distribution, indeed tend to restrict the possibilities of small-scale and large-amplitude EHs. But given the conditions of validity of the underlying gyrokinetic approach, our results can only show a modest effect for small scale EHs.

Since these existence conditions only determine the boundaries of a semi-open parameter space, our results (and previous ones, e.g., see discussion in Chen et al.\cite{chen2005width}) do not provide any strong constraints on the relationship between perpendicular and parallel lengthscales. Consequently, as suggested in the introduction, we believe that what determines the oblateness of these structures lies not so much in their self-consistent physics, but rather in the mechanisms by which they are generated. So, while we disagree with Hutchinson's approach\cite{hutchinson2021oblate} that the EH's positive core is shielded by passing electrons (rather than arising from their depletion due to their acceleration by the potential well; see discussion in section \ref{sec::electrondensities}) and thus that the electron polarization effect would virtually be an anisotropic shielding mechanism, we can agree with his conclusion given in the abstract that "trapped electron charge distribution anisotropy must [...] underlie the oblate shape of electron holes".

As the validity of the guiding center approximation for modelling EHs has been questionned, we would like to point out that the parallel magnetic field perturbation reported for fast EHs observed in the Earth's magnetotail, has so far been well explained by the $\mathbf{\delta E}\times \mathbf{B}_0$ electron drift current inside the hole.\cite{andersson2009new, tao2011model, holmes2018electron, steinvall2019observations}

Particle measurements by the MMS mission, even provided at 30 ms time resolution,\cite{burch2016electron} do not allow to fully resolve the electron distribution function in fast EHs, which have characteristic times of 1 ms. Depending on the number of EHs detected in 30 ms, measured electron distribution functions can be statistically representative to the EH or to the ambient plasma. Therefore, theoretical  developments are  crucial to better understand in situ observations of such self consistent structures. This could be addressed by PIC simulations and observations in weakly magnetized media ($\omega_c/\omega_p \lesssim 1$), in a forthcoming study.

\section*{Data Availability Statement}
Data sharing is not applicable to this article as no new data were created or analyzed in this
study. 


\providecommand{\noopsort}[1]{}\providecommand{\singleletter}[1]{#1}%
\begin{thebibliography}{51}%
\makeatletter
\providecommand \@ifxundefined [1]{%
 \@ifx{#1\undefined}
}%
\providecommand \@ifnum [1]{%
 \ifnum #1\expandafter \@firstoftwo
 \else \expandafter \@secondoftwo
 \fi
}%
\providecommand \@ifx [1]{%
 \ifx #1\expandafter \@firstoftwo
 \else \expandafter \@secondoftwo
 \fi
}%
\providecommand \natexlab [1]{#1}%
\providecommand \enquote  [1]{``#1''}%
\providecommand \bibnamefont  [1]{#1}%
\providecommand \bibfnamefont [1]{#1}%
\providecommand \citenamefont [1]{#1}%
\providecommand \href@noop [0]{\@secondoftwo}%
\providecommand \href [0]{\begingroup \@sanitize@url \@href}%
\providecommand \@href[1]{\@@startlink{#1}\@@href}%
\providecommand \@@href[1]{\endgroup#1\@@endlink}%
\providecommand \@sanitize@url [0]{\catcode `\\12\catcode `\$12\catcode
  `\&12\catcode `\#12\catcode `\^12\catcode `\_12\catcode `\%12\relax}%
\providecommand \@@startlink[1]{}%
\providecommand \@@endlink[0]{}%
\providecommand \url  [0]{\begingroup\@sanitize@url \@url }%
\providecommand \@url [1]{\endgroup\@href {#1}{\urlprefix }}%
\providecommand \urlprefix  [0]{URL }%
\providecommand \Eprint [0]{\href }%
\providecommand \doibase [0]{http://dx.doi.org/}%
\providecommand \selectlanguage [0]{\@gobble}%
\providecommand \bibinfo  [0]{\@secondoftwo}%
\providecommand \bibfield  [0]{\@secondoftwo}%
\providecommand \translation [1]{[#1]}%
\providecommand \BibitemOpen [0]{}%
\providecommand \bibitemStop [0]{}%
\providecommand \bibitemNoStop [0]{.\EOS\space}%
\providecommand \EOS [0]{\spacefactor3000\relax}%
\providecommand \BibitemShut  [1]{\csname bibitem#1\endcsname}%
\let\auto@bib@innerbib\@empty
\bibitem [{\citenamefont {Chen}, \citenamefont {Thouless},\ and\ \citenamefont
  {Tang}(2004)}]{chen2004bernstein}%
  \BibitemOpen
  \bibfield  {author} {\bibinfo {author} {\bibfnamefont {L.-J.}\ \bibnamefont
  {Chen}}, \bibinfo {author} {\bibfnamefont {D.~J.}\ \bibnamefont {Thouless}},
  \ and\ \bibinfo {author} {\bibfnamefont {J.-M.}\ \bibnamefont {Tang}},\
  }\bibfield  {title} {\enquote {\bibinfo {title} {Bernstein-greene-kruskal
  solitary waves in three-dimensional magnetized plasma},}\ }\href {\doibase
  10.1103/PhysRevE.69.055401} {\bibfield  {journal} {\bibinfo  {journal}
  {Physical Review E}\ }\textbf {\bibinfo {volume} {69}},\ \bibinfo {pages}
  {055401} (\bibinfo {year} {2004})}\BibitemShut {NoStop}%
\bibitem [{\citenamefont {Chen}\ \emph {et~al.}(2005)\citenamefont {Chen},
  \citenamefont {Pickett}, \citenamefont {Kintner}, \citenamefont {Franz},\
  and\ \citenamefont {Gurnett}}]{chen2005width}%
  \BibitemOpen
  \bibfield  {author} {\bibinfo {author} {\bibfnamefont {L.-J.}\ \bibnamefont
  {Chen}}, \bibinfo {author} {\bibfnamefont {J.}~\bibnamefont {Pickett}},
  \bibinfo {author} {\bibfnamefont {P.}~\bibnamefont {Kintner}}, \bibinfo
  {author} {\bibfnamefont {J.}~\bibnamefont {Franz}}, \ and\ \bibinfo {author}
  {\bibfnamefont {D.}~\bibnamefont {Gurnett}},\ }\bibfield  {title} {\enquote
  {\bibinfo {title} {On the width-amplitude inequality of electron phase space
  holes},}\ }\href {\doibase 10.1029/2005JA011087} {\bibfield  {journal}
  {\bibinfo  {journal} {J. Geophys. Res. Space Phys.}\ }\textbf {\bibinfo
  {volume} {110}} (\bibinfo {year} {2005}),\ 10.1029/2005JA011087}\BibitemShut
  {NoStop}%
\bibitem [{\citenamefont {Matsumoto}\ \emph {et~al.}(1994)\citenamefont
  {Matsumoto}, \citenamefont {Kojima}, \citenamefont {Miyatake}, \citenamefont
  {Omura}, \citenamefont {Okada}, \citenamefont {Nagano},\ and\ \citenamefont
  {Tsutsui}}]{matsumoto1994electrostatic}%
  \BibitemOpen
  \bibfield  {author} {\bibinfo {author} {\bibfnamefont {H.}~\bibnamefont
  {Matsumoto}}, \bibinfo {author} {\bibfnamefont {H.}~\bibnamefont {Kojima}},
  \bibinfo {author} {\bibfnamefont {T.}~\bibnamefont {Miyatake}}, \bibinfo
  {author} {\bibfnamefont {Y.}~\bibnamefont {Omura}}, \bibinfo {author}
  {\bibfnamefont {M.}~\bibnamefont {Okada}}, \bibinfo {author} {\bibfnamefont
  {I.}~\bibnamefont {Nagano}}, \ and\ \bibinfo {author} {\bibfnamefont
  {M.}~\bibnamefont {Tsutsui}},\ }\bibfield  {title} {\enquote {\bibinfo
  {title} {Electrostatic solitary waves (esw) in the magnetotail: Ben wave
  forms observed by geotail},}\ }\href {\doibase 10.1029/94GL01284} {\bibfield
  {journal} {\bibinfo  {journal} {Geophysical Research Letters}\ }\textbf
  {\bibinfo {volume} {21}},\ \bibinfo {pages} {2915--2918} (\bibinfo {year}
  {1994})}\BibitemShut {NoStop}%
\bibitem [{\citenamefont {Ergun}\ \emph {et~al.}(1998)\citenamefont {Ergun},
  \citenamefont {Carlson}, \citenamefont {McFadden}, \citenamefont {Mozer},
  \citenamefont {Delory}, \citenamefont {Peria}, \citenamefont {Chaston},
  \citenamefont {Temerin}, \citenamefont {Roth}, \citenamefont {Muschietti}
  \emph {et~al.}}]{ergun1998fast}%
  \BibitemOpen
  \bibfield  {author} {\bibinfo {author} {\bibfnamefont {R.}~\bibnamefont
  {Ergun}}, \bibinfo {author} {\bibfnamefont {C.}~\bibnamefont {Carlson}},
  \bibinfo {author} {\bibfnamefont {J.}~\bibnamefont {McFadden}}, \bibinfo
  {author} {\bibfnamefont {F.}~\bibnamefont {Mozer}}, \bibinfo {author}
  {\bibfnamefont {G.}~\bibnamefont {Delory}}, \bibinfo {author} {\bibfnamefont
  {W.}~\bibnamefont {Peria}}, \bibinfo {author} {\bibfnamefont
  {C.}~\bibnamefont {Chaston}}, \bibinfo {author} {\bibfnamefont
  {M.}~\bibnamefont {Temerin}}, \bibinfo {author} {\bibfnamefont
  {I.}~\bibnamefont {Roth}}, \bibinfo {author} {\bibfnamefont {L.}~\bibnamefont
  {Muschietti}},  \emph {et~al.},\ }\bibfield  {title} {\enquote {\bibinfo
  {title} {Fast satellite observations of large-amplitude solitary
  structures},}\ }\href {\doibase 10.1029/98GL00636} {\bibfield  {journal}
  {\bibinfo  {journal} {Geophys. Res. Lett}\ }\textbf {\bibinfo {volume}
  {25}},\ \bibinfo {pages} {2041--2044} (\bibinfo {year} {1998})}\BibitemShut
  {NoStop}%
\bibitem [{\citenamefont {Bale}\ \emph {et~al.}(1998)\citenamefont {Bale},
  \citenamefont {Kellogg}, \citenamefont {Larsen}, \citenamefont {Lin},
  \citenamefont {Goetz},\ and\ \citenamefont {Lepping}}]{bale1998bipolar}%
  \BibitemOpen
  \bibfield  {author} {\bibinfo {author} {\bibfnamefont {S.}~\bibnamefont
  {Bale}}, \bibinfo {author} {\bibfnamefont {P.}~\bibnamefont {Kellogg}},
  \bibinfo {author} {\bibfnamefont {D.}~\bibnamefont {Larsen}}, \bibinfo
  {author} {\bibfnamefont {R.}~\bibnamefont {Lin}}, \bibinfo {author}
  {\bibfnamefont {K.}~\bibnamefont {Goetz}}, \ and\ \bibinfo {author}
  {\bibfnamefont {R.}~\bibnamefont {Lepping}},\ }\bibfield  {title} {\enquote
  {\bibinfo {title} {Bipolar electrostatic structures in the shock transition
  region: Evidence of electron phase space holes},}\ }\href {\doibase
  10.1029/98GL02111} {\bibfield  {journal} {\bibinfo  {journal} {Geophys. Res.
  Lett}\ }\textbf {\bibinfo {volume} {25}},\ \bibinfo {pages} {2929--2932}
  (\bibinfo {year} {1998})}\BibitemShut {NoStop}%
\bibitem [{\citenamefont {Franz}\ \emph {et~al.}(2005)\citenamefont {Franz},
  \citenamefont {Kintner}, \citenamefont {Pickett},\ and\ \citenamefont
  {Chen}}]{franz2005properties}%
  \BibitemOpen
  \bibfield  {author} {\bibinfo {author} {\bibfnamefont {J.}~\bibnamefont
  {Franz}}, \bibinfo {author} {\bibfnamefont {P.}~\bibnamefont {Kintner}},
  \bibinfo {author} {\bibfnamefont {J.}~\bibnamefont {Pickett}}, \ and\
  \bibinfo {author} {\bibfnamefont {L.-J.}\ \bibnamefont {Chen}},\ }\bibfield
  {title} {\enquote {\bibinfo {title} {Properties of small-amplitude electron
  phase-space holes observed by polar},}\ }\href@noop {} {\bibfield  {journal}
  {\bibinfo  {journal} {Journal of Geophysical Research: Space Physics}\
  }\textbf {\bibinfo {volume} {110}} (\bibinfo {year} {2005})}\BibitemShut
  {NoStop}%
\bibitem [{\citenamefont {Cattell}\ \emph {et~al.}(2005)\citenamefont
  {Cattell}, \citenamefont {Dombeck}, \citenamefont {Wygant}, \citenamefont
  {Drake}, \citenamefont {Swisdak}, \citenamefont {Goldstein}, \citenamefont
  {Keith}, \citenamefont {Fazakerley}, \citenamefont {Andr{\'e}}, \citenamefont
  {Lucek} \emph {et~al.}}]{cattell2005cluster}%
  \BibitemOpen
  \bibfield  {author} {\bibinfo {author} {\bibfnamefont {C.}~\bibnamefont
  {Cattell}}, \bibinfo {author} {\bibfnamefont {J.}~\bibnamefont {Dombeck}},
  \bibinfo {author} {\bibfnamefont {J.}~\bibnamefont {Wygant}}, \bibinfo
  {author} {\bibfnamefont {J.}~\bibnamefont {Drake}}, \bibinfo {author}
  {\bibfnamefont {M.}~\bibnamefont {Swisdak}}, \bibinfo {author} {\bibfnamefont
  {M.}~\bibnamefont {Goldstein}}, \bibinfo {author} {\bibfnamefont
  {W.}~\bibnamefont {Keith}}, \bibinfo {author} {\bibfnamefont
  {A.}~\bibnamefont {Fazakerley}}, \bibinfo {author} {\bibfnamefont
  {M.}~\bibnamefont {Andr{\'e}}}, \bibinfo {author} {\bibfnamefont
  {E.}~\bibnamefont {Lucek}},  \emph {et~al.},\ }\bibfield  {title} {\enquote
  {\bibinfo {title} {Cluster observations of electron holes in association with
  magnetotail reconnection and comparison to simulations},}\ }\href {\doibase
  10.1029/2004JA010519} {\bibfield  {journal} {\bibinfo  {journal} {Journal of
  Geophysical Research: Space Physics}\ }\textbf {\bibinfo {volume} {110}}
  (\bibinfo {year} {2005}),\ 10.1029/2004JA010519}\BibitemShut {NoStop}%
\bibitem [{\citenamefont {Norgren}\ \emph {et~al.}(2015)\citenamefont
  {Norgren}, \citenamefont {Andr{\'e}}, \citenamefont {Vaivads},\ and\
  \citenamefont {Khotyaintsev}}]{norgren2015slow}%
  \BibitemOpen
  \bibfield  {author} {\bibinfo {author} {\bibfnamefont {C.}~\bibnamefont
  {Norgren}}, \bibinfo {author} {\bibfnamefont {M.}~\bibnamefont {Andr{\'e}}},
  \bibinfo {author} {\bibfnamefont {A.}~\bibnamefont {Vaivads}}, \ and\
  \bibinfo {author} {\bibfnamefont {Y.~V.}\ \bibnamefont {Khotyaintsev}},\
  }\bibfield  {title} {\enquote {\bibinfo {title} {Slow electron phase space
  holes: Magnetotail observations},}\ }\href {\doibase 10.1002/2015GL063218}
  {\bibfield  {journal} {\bibinfo  {journal} {Geophys. Res. Lett}\ }\textbf
  {\bibinfo {volume} {42}},\ \bibinfo {pages} {1654--1661} (\bibinfo {year}
  {2015})}\BibitemShut {NoStop}%
\bibitem [{\citenamefont {Fu}\ \emph {et~al.}(2020)\citenamefont {Fu},
  \citenamefont {Chen}, \citenamefont {Chen}, \citenamefont {Xu}, \citenamefont
  {Wang}, \citenamefont {Liu}, \citenamefont {Liu}, \citenamefont
  {Khotyaintsev}, \citenamefont {Ergun}, \citenamefont {Giles} \emph
  {et~al.}}]{fu2020first}%
  \BibitemOpen
  \bibfield  {author} {\bibinfo {author} {\bibfnamefont {H.}~\bibnamefont
  {Fu}}, \bibinfo {author} {\bibfnamefont {F.}~\bibnamefont {Chen}}, \bibinfo
  {author} {\bibfnamefont {Z.}~\bibnamefont {Chen}}, \bibinfo {author}
  {\bibfnamefont {Y.}~\bibnamefont {Xu}}, \bibinfo {author} {\bibfnamefont
  {Z.}~\bibnamefont {Wang}}, \bibinfo {author} {\bibfnamefont {Y.}~\bibnamefont
  {Liu}}, \bibinfo {author} {\bibfnamefont {C.}~\bibnamefont {Liu}}, \bibinfo
  {author} {\bibfnamefont {Y.~V.}\ \bibnamefont {Khotyaintsev}}, \bibinfo
  {author} {\bibfnamefont {R.}~\bibnamefont {Ergun}}, \bibinfo {author}
  {\bibfnamefont {B.}~\bibnamefont {Giles}},  \emph {et~al.},\ }\bibfield
  {title} {\enquote {\bibinfo {title} {First measurements of electrons and
  waves inside an electrostatic solitary wave},}\ }\href@noop {} {\bibfield
  {journal} {\bibinfo  {journal} {Physical review letters}\ }\textbf {\bibinfo
  {volume} {124}},\ \bibinfo {pages} {095101} (\bibinfo {year}
  {2020})}\BibitemShut {NoStop}%
\bibitem [{\citenamefont {Le~Contel}\ \emph {et~al.}(2017)\citenamefont
  {Le~Contel}, \citenamefont {Nakamura}, \citenamefont {Breuillard},
  \citenamefont {Argall}, \citenamefont {Graham}, \citenamefont {Fischer},
  \citenamefont {Retin{\`o}}, \citenamefont {Berthomier}, \citenamefont
  {Pottelette}, \citenamefont {Mirioni} \emph {et~al.}}]{le2017lower}%
  \BibitemOpen
  \bibfield  {author} {\bibinfo {author} {\bibfnamefont {O.}~\bibnamefont
  {Le~Contel}}, \bibinfo {author} {\bibfnamefont {R.}~\bibnamefont {Nakamura}},
  \bibinfo {author} {\bibfnamefont {H.}~\bibnamefont {Breuillard}}, \bibinfo
  {author} {\bibfnamefont {M.}~\bibnamefont {Argall}}, \bibinfo {author}
  {\bibfnamefont {D.~B.}\ \bibnamefont {Graham}}, \bibinfo {author}
  {\bibfnamefont {D.}~\bibnamefont {Fischer}}, \bibinfo {author} {\bibfnamefont
  {A.}~\bibnamefont {Retin{\`o}}}, \bibinfo {author} {\bibfnamefont
  {M.}~\bibnamefont {Berthomier}}, \bibinfo {author} {\bibfnamefont
  {R.}~\bibnamefont {Pottelette}}, \bibinfo {author} {\bibfnamefont
  {L.}~\bibnamefont {Mirioni}},  \emph {et~al.},\ }\bibfield  {title} {\enquote
  {\bibinfo {title} {Lower hybrid drift waves and electromagnetic electron
  space-phase holes associated with dipolarization fronts and field-aligned
  currents observed by the magnetospheric multiscale mission during a
  substorm},}\ }\href {\doibase 10.1002/2017JA024550} {\bibfield  {journal}
  {\bibinfo  {journal} {J. Geophys. Res. Space Phys.}\ }\textbf {\bibinfo
  {volume} {122}},\ \bibinfo {pages} {12--236} (\bibinfo {year}
  {2017})}\BibitemShut {NoStop}%
\bibitem [{\citenamefont {Tong}\ \emph {et~al.}(2018)\citenamefont {Tong},
  \citenamefont {Vasko}, \citenamefont {Mozer}, \citenamefont {Bale},
  \citenamefont {Roth}, \citenamefont {Artemyev}, \citenamefont {Ergun},
  \citenamefont {Giles}, \citenamefont {Lindqvist}, \citenamefont {Russell}
  \emph {et~al.}}]{tong2018simultaneous}%
  \BibitemOpen
  \bibfield  {author} {\bibinfo {author} {\bibfnamefont {Y.}~\bibnamefont
  {Tong}}, \bibinfo {author} {\bibfnamefont {I.}~\bibnamefont {Vasko}},
  \bibinfo {author} {\bibfnamefont {F.}~\bibnamefont {Mozer}}, \bibinfo
  {author} {\bibfnamefont {S.~D.}\ \bibnamefont {Bale}}, \bibinfo {author}
  {\bibfnamefont {I.}~\bibnamefont {Roth}}, \bibinfo {author} {\bibfnamefont
  {A.}~\bibnamefont {Artemyev}}, \bibinfo {author} {\bibfnamefont
  {R.}~\bibnamefont {Ergun}}, \bibinfo {author} {\bibfnamefont
  {B.}~\bibnamefont {Giles}}, \bibinfo {author} {\bibfnamefont {P.-A.}\
  \bibnamefont {Lindqvist}}, \bibinfo {author} {\bibfnamefont {C.}~\bibnamefont
  {Russell}},  \emph {et~al.},\ }\bibfield  {title} {\enquote {\bibinfo {title}
  {Simultaneous multispacecraft probing of electron phase space holes},}\
  }\href {\doibase 10.1029/2018GL079044} {\bibfield  {journal} {\bibinfo
  {journal} {Geophys. Res. Lett}\ }\textbf {\bibinfo {volume} {45}},\ \bibinfo
  {pages} {11--513} (\bibinfo {year} {2018})}\BibitemShut {NoStop}%
\bibitem [{\citenamefont {Holmes}\ \emph {et~al.}(2018)\citenamefont {Holmes},
  \citenamefont {Ergun}, \citenamefont {Newman}, \citenamefont {Ahmadi},
  \citenamefont {Andersson}, \citenamefont {Le~Contel}, \citenamefont
  {Torbert}, \citenamefont {Giles}, \citenamefont {Strangeway},\ and\
  \citenamefont {Burch}}]{holmes2018electron}%
  \BibitemOpen
  \bibfield  {author} {\bibinfo {author} {\bibfnamefont {J.}~\bibnamefont
  {Holmes}}, \bibinfo {author} {\bibfnamefont {R.}~\bibnamefont {Ergun}},
  \bibinfo {author} {\bibfnamefont {D.}~\bibnamefont {Newman}}, \bibinfo
  {author} {\bibfnamefont {N.}~\bibnamefont {Ahmadi}}, \bibinfo {author}
  {\bibfnamefont {L.}~\bibnamefont {Andersson}}, \bibinfo {author}
  {\bibfnamefont {O.}~\bibnamefont {Le~Contel}}, \bibinfo {author}
  {\bibfnamefont {R.}~\bibnamefont {Torbert}}, \bibinfo {author} {\bibfnamefont
  {B.}~\bibnamefont {Giles}}, \bibinfo {author} {\bibfnamefont
  {R.}~\bibnamefont {Strangeway}}, \ and\ \bibinfo {author} {\bibfnamefont
  {J.}~\bibnamefont {Burch}},\ }\bibfield  {title} {\enquote {\bibinfo {title}
  {Electron phase-space holes in three dimensions: Multispacecraft observations
  by magnetospheric multiscale},}\ }\href {\doibase 10.1029/2018JA025750}
  {\bibfield  {journal} {\bibinfo  {journal} {J. Geophys. Res. Space Phys.}\
  }\textbf {\bibinfo {volume} {123}},\ \bibinfo {pages} {9963--9978} (\bibinfo
  {year} {2018})}\BibitemShut {NoStop}%
\bibitem [{\citenamefont {Steinvall}\ \emph {et~al.}(2019)\citenamefont
  {Steinvall}, \citenamefont {Khotyaintsev}, \citenamefont {Graham},
  \citenamefont {Vaivads}, \citenamefont {Le~Contel},\ and\ \citenamefont
  {Russell}}]{steinvall2019observations}%
  \BibitemOpen
  \bibfield  {author} {\bibinfo {author} {\bibfnamefont {K.}~\bibnamefont
  {Steinvall}}, \bibinfo {author} {\bibfnamefont {Y.~V.}\ \bibnamefont
  {Khotyaintsev}}, \bibinfo {author} {\bibfnamefont {D.~B.}\ \bibnamefont
  {Graham}}, \bibinfo {author} {\bibfnamefont {A.}~\bibnamefont {Vaivads}},
  \bibinfo {author} {\bibfnamefont {O.}~\bibnamefont {Le~Contel}}, \ and\
  \bibinfo {author} {\bibfnamefont {C.~T.}\ \bibnamefont {Russell}},\
  }\bibfield  {title} {\enquote {\bibinfo {title} {Observations of
  electromagnetic electron holes and evidence of cherenkov whistler
  emission},}\ }\href {\doibase 10.1103/PhysRevLett.123.255101} {\bibfield
  {journal} {\bibinfo  {journal} {Phys. Rev. Lett}\ }\textbf {\bibinfo {volume}
  {123}},\ \bibinfo {pages} {255101} (\bibinfo {year} {2019})}\BibitemShut
  {NoStop}%
\bibitem [{\citenamefont {Andersson}\ \emph {et~al.}(2009)\citenamefont
  {Andersson}, \citenamefont {Ergun}, \citenamefont {Tao}, \citenamefont
  {Roux}, \citenamefont {Le~Contel}, \citenamefont {Angelopoulos},
  \citenamefont {Bonnell}, \citenamefont {McFadden}, \citenamefont {Larson},
  \citenamefont {Eriksson} \emph {et~al.}}]{andersson2009new}%
  \BibitemOpen
  \bibfield  {author} {\bibinfo {author} {\bibfnamefont {L.}~\bibnamefont
  {Andersson}}, \bibinfo {author} {\bibfnamefont {R.}~\bibnamefont {Ergun}},
  \bibinfo {author} {\bibfnamefont {J.}~\bibnamefont {Tao}}, \bibinfo {author}
  {\bibfnamefont {A.}~\bibnamefont {Roux}}, \bibinfo {author} {\bibfnamefont
  {O.}~\bibnamefont {Le~Contel}}, \bibinfo {author} {\bibfnamefont
  {V.}~\bibnamefont {Angelopoulos}}, \bibinfo {author} {\bibfnamefont
  {J.}~\bibnamefont {Bonnell}}, \bibinfo {author} {\bibfnamefont
  {J.}~\bibnamefont {McFadden}}, \bibinfo {author} {\bibfnamefont
  {D.}~\bibnamefont {Larson}}, \bibinfo {author} {\bibfnamefont
  {S.}~\bibnamefont {Eriksson}},  \emph {et~al.},\ }\bibfield  {title}
  {\enquote {\bibinfo {title} {New features of electron phase space holes
  observed by the themis mission},}\ }\href {\doibase
  10.1103/PhysRevLett.102.225004} {\bibfield  {journal} {\bibinfo  {journal}
  {Phys. Rev. Lett}\ }\textbf {\bibinfo {volume} {102}},\ \bibinfo {pages}
  {225004} (\bibinfo {year} {2009})}\BibitemShut {NoStop}%
\bibitem [{\citenamefont {Tao}\ \emph {et~al.}(2011)\citenamefont {Tao},
  \citenamefont {Ergun}, \citenamefont {Andersson}, \citenamefont {Bonnell},
  \citenamefont {Roux}, \citenamefont {Le~Contel}, \citenamefont
  {Angelopoulos}, \citenamefont {McFadden}, \citenamefont {Larson},
  \citenamefont {Cully} \emph {et~al.}}]{tao2011model}%
  \BibitemOpen
  \bibfield  {author} {\bibinfo {author} {\bibfnamefont {J.}~\bibnamefont
  {Tao}}, \bibinfo {author} {\bibfnamefont {R.}~\bibnamefont {Ergun}}, \bibinfo
  {author} {\bibfnamefont {L.}~\bibnamefont {Andersson}}, \bibinfo {author}
  {\bibfnamefont {J.}~\bibnamefont {Bonnell}}, \bibinfo {author} {\bibfnamefont
  {A.}~\bibnamefont {Roux}}, \bibinfo {author} {\bibfnamefont {O.}~\bibnamefont
  {Le~Contel}}, \bibinfo {author} {\bibfnamefont {V.}~\bibnamefont
  {Angelopoulos}}, \bibinfo {author} {\bibfnamefont {J.}~\bibnamefont
  {McFadden}}, \bibinfo {author} {\bibfnamefont {D.}~\bibnamefont {Larson}},
  \bibinfo {author} {\bibfnamefont {C.~M.}\ \bibnamefont {Cully}},  \emph
  {et~al.},\ }\bibfield  {title} {\enquote {\bibinfo {title} {A model of
  electromagnetic electron phase-space holes and its application},}\ }\href
  {\doibase 10.1029/2010JA016054} {\bibfield  {journal} {\bibinfo  {journal}
  {J. Geophys. Res. Space Phys.}\ }\textbf {\bibinfo {volume} {116}} (\bibinfo
  {year} {2011}),\ 10.1029/2010JA016054}\BibitemShut {NoStop}%
\bibitem [{\citenamefont {Shustov}\ \emph {et~al.}(2021)\citenamefont
  {Shustov}, \citenamefont {Kuzichev}, \citenamefont {Vasko}, \citenamefont
  {Artemyev},\ and\ \citenamefont {Gerrard}}]{shustov2021dynamics}%
  \BibitemOpen
  \bibfield  {author} {\bibinfo {author} {\bibfnamefont {P.~I.}\ \bibnamefont
  {Shustov}}, \bibinfo {author} {\bibfnamefont {I.~V.}\ \bibnamefont
  {Kuzichev}}, \bibinfo {author} {\bibfnamefont {I.~Y.}\ \bibnamefont {Vasko}},
  \bibinfo {author} {\bibfnamefont {A.~V.}\ \bibnamefont {Artemyev}}, \ and\
  \bibinfo {author} {\bibfnamefont {A.~J.}\ \bibnamefont {Gerrard}},\
  }\bibfield  {title} {\enquote {\bibinfo {title} {The dynamics of electron
  holes in current sheets},}\ }\href {\doibase 10.1063/5.0029999} {\bibfield
  {journal} {\bibinfo  {journal} {Phys. Plasmas}\ }\textbf {\bibinfo {volume}
  {28}},\ \bibinfo {pages} {012902} (\bibinfo {year} {2021})}\BibitemShut
  {NoStop}%
\bibitem [{\citenamefont {Williams}\ \emph {et~al.}(2006)\citenamefont
  {Williams}, \citenamefont {Chen}, \citenamefont {Kurth}, \citenamefont
  {Gurnett},\ and\ \citenamefont {Dougherty}}]{williams2006electrostatic}%
  \BibitemOpen
  \bibfield  {author} {\bibinfo {author} {\bibfnamefont {J.}~\bibnamefont
  {Williams}}, \bibinfo {author} {\bibfnamefont {L.-J.}\ \bibnamefont {Chen}},
  \bibinfo {author} {\bibfnamefont {W.}~\bibnamefont {Kurth}}, \bibinfo
  {author} {\bibfnamefont {D.}~\bibnamefont {Gurnett}}, \ and\ \bibinfo
  {author} {\bibfnamefont {M.}~\bibnamefont {Dougherty}},\ }\bibfield  {title}
  {\enquote {\bibinfo {title} {Electrostatic solitary structures observed at
  saturn},}\ }\href {\doibase 10.1029/2005GL024532} {\bibfield  {journal}
  {\bibinfo  {journal} {Geophys. Res. Lett}\ }\textbf {\bibinfo {volume} {33}}
  (\bibinfo {year} {2006}),\ 10.1029/2005GL024532}\BibitemShut {NoStop}%
\bibitem [{\citenamefont {Pickett}\ \emph {et~al.}(2015)\citenamefont
  {Pickett}, \citenamefont {Kurth}, \citenamefont {Gurnett}, \citenamefont
  {Huff}, \citenamefont {Faden}, \citenamefont {Averkamp}, \citenamefont
  {P{\'\i}{\v{s}}a},\ and\ \citenamefont {Jones}}]{pickett2015electrostatic}%
  \BibitemOpen
  \bibfield  {author} {\bibinfo {author} {\bibfnamefont {J.}~\bibnamefont
  {Pickett}}, \bibinfo {author} {\bibfnamefont {W.}~\bibnamefont {Kurth}},
  \bibinfo {author} {\bibfnamefont {D.}~\bibnamefont {Gurnett}}, \bibinfo
  {author} {\bibfnamefont {R.}~\bibnamefont {Huff}}, \bibinfo {author}
  {\bibfnamefont {J.}~\bibnamefont {Faden}}, \bibinfo {author} {\bibfnamefont
  {T.}~\bibnamefont {Averkamp}}, \bibinfo {author} {\bibfnamefont
  {D.}~\bibnamefont {P{\'\i}{\v{s}}a}}, \ and\ \bibinfo {author} {\bibfnamefont
  {G.}~\bibnamefont {Jones}},\ }\bibfield  {title} {\enquote {\bibinfo {title}
  {Electrostatic solitary waves observed at saturn by cassini inside 10 rs and
  near enceladus},}\ }\href {\doibase 10.1002/2015JA021305} {\bibfield
  {journal} {\bibinfo  {journal} {J. Geophys. Res. Space Phys.}\ }\textbf
  {\bibinfo {volume} {120}},\ \bibinfo {pages} {6569--6580} (\bibinfo {year}
  {2015})}\BibitemShut {NoStop}%
\bibitem [{\citenamefont {Malaspina}\ \emph {et~al.}(2020)\citenamefont
  {Malaspina}, \citenamefont {Goodrich}, \citenamefont {Livi}, \citenamefont
  {Halekas}, \citenamefont {McManus}, \citenamefont {Curry}, \citenamefont
  {Bale}, \citenamefont {Bonnell}, \citenamefont {de~Wit}, \citenamefont
  {Goetz} \emph {et~al.}}]{malaspina2020plasma}%
  \BibitemOpen
  \bibfield  {author} {\bibinfo {author} {\bibfnamefont {D.~M.}\ \bibnamefont
  {Malaspina}}, \bibinfo {author} {\bibfnamefont {K.}~\bibnamefont {Goodrich}},
  \bibinfo {author} {\bibfnamefont {R.}~\bibnamefont {Livi}}, \bibinfo {author}
  {\bibfnamefont {J.}~\bibnamefont {Halekas}}, \bibinfo {author} {\bibfnamefont
  {M.}~\bibnamefont {McManus}}, \bibinfo {author} {\bibfnamefont
  {S.}~\bibnamefont {Curry}}, \bibinfo {author} {\bibfnamefont {S.~D.}\
  \bibnamefont {Bale}}, \bibinfo {author} {\bibfnamefont {J.~W.}\ \bibnamefont
  {Bonnell}}, \bibinfo {author} {\bibfnamefont {T.~D.}\ \bibnamefont {de~Wit}},
  \bibinfo {author} {\bibfnamefont {K.}~\bibnamefont {Goetz}},  \emph
  {et~al.},\ }\bibfield  {title} {\enquote {\bibinfo {title} {Plasma double
  layers at the boundary between venus and the solar wind},}\ }\href {\doibase
  10.1029/2020GL090115} {\bibfield  {journal} {\bibinfo  {journal} {Geophys.
  Res. Lett}\ }\textbf {\bibinfo {volume} {47}},\ \bibinfo {pages}
  {e2020GL090115} (\bibinfo {year} {2020})}\BibitemShut {NoStop}%
\bibitem [{\citenamefont {Hadid}\ \emph {et~al.}(2021)\citenamefont {Hadid},
  \citenamefont {Edberg}, \citenamefont {Chust}, \citenamefont
  {P{\'\i}{\v{s}}a}, \citenamefont {Dimmock}, \citenamefont {Morooka},
  \citenamefont {Maksimovic}, \citenamefont {Khotyaintsev}, \citenamefont
  {Sou{\v{c}}ek}, \citenamefont {Kretzschmar}, \citenamefont {Vecchio},
  \citenamefont {Le~Contel}, \citenamefont {Retin{\'o}}, \citenamefont {Allen},
  \citenamefont {Volwerk}, \citenamefont {Fowler}, \citenamefont
  {Sorriso-Valvo}, \citenamefont {Karlsson},\ and\ \citenamefont
  {et~al.}}]{haddid2021Solar}%
  \BibitemOpen
  \bibfield  {author} {\bibinfo {author} {\bibfnamefont {L.}~\bibnamefont
  {Hadid}}, \bibinfo {author} {\bibfnamefont {N.}~\bibnamefont {Edberg}},
  \bibinfo {author} {\bibfnamefont {T.}~\bibnamefont {Chust}}, \bibinfo
  {author} {\bibfnamefont {D.}~\bibnamefont {P{\'\i}{\v{s}}a}}, \bibinfo
  {author} {\bibfnamefont {A.}~\bibnamefont {Dimmock}}, \bibinfo {author}
  {\bibfnamefont {M.}~\bibnamefont {Morooka}}, \bibinfo {author} {\bibfnamefont
  {M.}~\bibnamefont {Maksimovic}}, \bibinfo {author} {\bibfnamefont
  {Y.}~\bibnamefont {Khotyaintsev}}, \bibinfo {author} {\bibfnamefont
  {J.}~\bibnamefont {Sou{\v{c}}ek}}, \bibinfo {author} {\bibfnamefont
  {M.}~\bibnamefont {Kretzschmar}}, \bibinfo {author} {\bibfnamefont
  {A.}~\bibnamefont {Vecchio}}, \bibinfo {author} {\bibfnamefont
  {O.}~\bibnamefont {Le~Contel}}, \bibinfo {author} {\bibfnamefont
  {A.}~\bibnamefont {Retin{\'o}}}, \bibinfo {author} {\bibfnamefont
  {R.}~\bibnamefont {Allen}}, \bibinfo {author} {\bibfnamefont
  {M.}~\bibnamefont {Volwerk}}, \bibinfo {author} {\bibfnamefont
  {C.}~\bibnamefont {Fowler}}, \bibinfo {author} {\bibfnamefont
  {L.}~\bibnamefont {Sorriso-Valvo}}, \bibinfo {author} {\bibfnamefont
  {T.}~\bibnamefont {Karlsson}}, \ and\ \bibinfo {author} {\bibnamefont
  {et~al.}},\ }\bibfield  {title} {\enquote {\bibinfo {title} {Solar
  orbiter’s first venus flyby: observations from the radio and plasma wave
  instrument},}\ }\href@noop {} {\bibfield  {journal} {\bibinfo  {journal}
  {Submitted to Astron. Astrophys.}\ } (\bibinfo {year} {2021})}\BibitemShut
  {NoStop}%
\bibitem [{\citenamefont {Montgomery}\ \emph {et~al.}(2001)\citenamefont
  {Montgomery}, \citenamefont {Focia}, \citenamefont {Rose}, \citenamefont
  {Russell}, \citenamefont {Cobble}, \citenamefont {Fern{\'a}ndez},\ and\
  \citenamefont {Johnson}}]{montgomery2001observation}%
  \BibitemOpen
  \bibfield  {author} {\bibinfo {author} {\bibfnamefont {D.}~\bibnamefont
  {Montgomery}}, \bibinfo {author} {\bibfnamefont {R.}~\bibnamefont {Focia}},
  \bibinfo {author} {\bibfnamefont {H.}~\bibnamefont {Rose}}, \bibinfo {author}
  {\bibfnamefont {D.}~\bibnamefont {Russell}}, \bibinfo {author} {\bibfnamefont
  {J.}~\bibnamefont {Cobble}}, \bibinfo {author} {\bibfnamefont
  {J.}~\bibnamefont {Fern{\'a}ndez}}, \ and\ \bibinfo {author} {\bibfnamefont
  {R.}~\bibnamefont {Johnson}},\ }\bibfield  {title} {\enquote {\bibinfo
  {title} {Observation of stimulated electron-acoustic-wave scattering},}\
  }\href {\doibase 10.1103/PhysRevLett.87.155001} {\bibfield  {journal}
  {\bibinfo  {journal} {Phys. Rev. lett}\ }\textbf {\bibinfo {volume} {87}},\
  \bibinfo {pages} {155001} (\bibinfo {year} {2001})}\BibitemShut {NoStop}%
\bibitem [{\citenamefont {Fox}\ \emph {et~al.}(2008)\citenamefont {Fox},
  \citenamefont {Porkolab}, \citenamefont {Egedal}, \citenamefont {Katz},\ and\
  \citenamefont {Le}}]{fox2008laboratory}%
  \BibitemOpen
  \bibfield  {author} {\bibinfo {author} {\bibfnamefont {W.}~\bibnamefont
  {Fox}}, \bibinfo {author} {\bibfnamefont {M.}~\bibnamefont {Porkolab}},
  \bibinfo {author} {\bibfnamefont {J.}~\bibnamefont {Egedal}}, \bibinfo
  {author} {\bibfnamefont {N.}~\bibnamefont {Katz}}, \ and\ \bibinfo {author}
  {\bibfnamefont {A.}~\bibnamefont {Le}},\ }\bibfield  {title} {\enquote
  {\bibinfo {title} {Laboratory observation of electron phase-space holes
  during magnetic reconnection},}\ }\href {\doibase
  10.1103/PhysRevLett.101.255003} {\bibfield  {journal} {\bibinfo  {journal}
  {Phys. Rev. Lett}\ }\textbf {\bibinfo {volume} {101}},\ \bibinfo {pages}
  {255003} (\bibinfo {year} {2008})}\BibitemShut {NoStop}%
\bibitem [{\citenamefont {Lefebvre}\ \emph {et~al.}(2010)\citenamefont
  {Lefebvre}, \citenamefont {Chen}, \citenamefont {Gekelman}, \citenamefont
  {Kintner}, \citenamefont {Pickett}, \citenamefont {Pribyl}, \citenamefont
  {Vincena}, \citenamefont {Chiang},\ and\ \citenamefont
  {Judy}}]{lefebvre2010laboratory}%
  \BibitemOpen
  \bibfield  {author} {\bibinfo {author} {\bibfnamefont {B.}~\bibnamefont
  {Lefebvre}}, \bibinfo {author} {\bibfnamefont {L.-J.}\ \bibnamefont {Chen}},
  \bibinfo {author} {\bibfnamefont {W.}~\bibnamefont {Gekelman}}, \bibinfo
  {author} {\bibfnamefont {P.}~\bibnamefont {Kintner}}, \bibinfo {author}
  {\bibfnamefont {J.}~\bibnamefont {Pickett}}, \bibinfo {author} {\bibfnamefont
  {P.}~\bibnamefont {Pribyl}}, \bibinfo {author} {\bibfnamefont
  {S.}~\bibnamefont {Vincena}}, \bibinfo {author} {\bibfnamefont
  {F.}~\bibnamefont {Chiang}}, \ and\ \bibinfo {author} {\bibfnamefont
  {J.}~\bibnamefont {Judy}},\ }\bibfield  {title} {\enquote {\bibinfo {title}
  {Laboratory measurements of electrostatic solitary structures generated by
  beam injection},}\ }\href {\doibase 10.1103/PhysRevLett.105.115001}
  {\bibfield  {journal} {\bibinfo  {journal} {Phys. Rev Lett}\ }\textbf
  {\bibinfo {volume} {105}},\ \bibinfo {pages} {115001} (\bibinfo {year}
  {2010})}\BibitemShut {NoStop}%
\bibitem [{\citenamefont {Mamun}\ and\ \citenamefont
  {Shukla}(2010)}]{mamun2010solitary}%
  \BibitemOpen
  \bibfield  {author} {\bibinfo {author} {\bibfnamefont {A.}~\bibnamefont
  {Mamun}}\ and\ \bibinfo {author} {\bibfnamefont {P.}~\bibnamefont {Shukla}},\
  }\bibfield  {title} {\enquote {\bibinfo {title} {Solitary waves in an
  ultrarelativistic degenerate dense plasma},}\ }\href {\doibase
  10.1063/1.3491433} {\bibfield  {journal} {\bibinfo  {journal} {Phys.
  Plasmas}\ }\textbf {\bibinfo {volume} {17}},\ \bibinfo {pages} {104504}
  (\bibinfo {year} {2010})}\BibitemShut {NoStop}%
\bibitem [{\citenamefont {Bernstein}, \citenamefont {Greene},\ and\
  \citenamefont {Kruskal}(1957)}]{bernstein1957exact}%
  \BibitemOpen
  \bibfield  {author} {\bibinfo {author} {\bibfnamefont {I.~B.}\ \bibnamefont
  {Bernstein}}, \bibinfo {author} {\bibfnamefont {J.~M.}\ \bibnamefont
  {Greene}}, \ and\ \bibinfo {author} {\bibfnamefont {M.~D.}\ \bibnamefont
  {Kruskal}},\ }\bibfield  {title} {\enquote {\bibinfo {title} {Exact nonlinear
  plasma oscillations},}\ }\href {\doibase 10.1103/PhysRev.108.546} {\bibfield
  {journal} {\bibinfo  {journal} {Phys. Rev.}\ }\textbf {\bibinfo {volume}
  {108}},\ \bibinfo {pages} {546} (\bibinfo {year} {1957})}\BibitemShut
  {NoStop}%
\bibitem [{\citenamefont {Haas}(2020)}]{haas2020bernstein}%
  \BibitemOpen
  \bibfield  {author} {\bibinfo {author} {\bibfnamefont {F.}~\bibnamefont
  {Haas}},\ }\bibfield  {title} {\enquote {\bibinfo {title}
  {Bernstein-greene-kruskal approach for the quantum vlasov equation},}\ }\href
  {\doibase 10.1209/0295-5075/132/20006} {\bibfield  {journal} {\bibinfo
  {journal} {EPL}\ }\textbf {\bibinfo {volume} {132}},\ \bibinfo {pages}
  {20006} (\bibinfo {year} {2020})}\BibitemShut {NoStop}%
\bibitem [{\citenamefont {Dauxois}\ and\ \citenamefont
  {Peyrard}(2006)}]{dauxois2006physics}%
  \BibitemOpen
  \bibfield  {author} {\bibinfo {author} {\bibfnamefont {T.}~\bibnamefont
  {Dauxois}}\ and\ \bibinfo {author} {\bibfnamefont {M.}~\bibnamefont
  {Peyrard}},\ }\href@noop {} {\emph {\bibinfo {title} {Physics of solitons}}}\
  (\bibinfo  {publisher} {Cambridge University Press},\ \bibinfo {year}
  {2006})\BibitemShut {NoStop}%
\bibitem [{\citenamefont {Krasovsky}, \citenamefont {Matsumoto},\ and\
  \citenamefont {Omura}(1997)}]{krasovsky1997bernstein}%
  \BibitemOpen
  \bibfield  {author} {\bibinfo {author} {\bibfnamefont {V.}~\bibnamefont
  {Krasovsky}}, \bibinfo {author} {\bibfnamefont {H.}~\bibnamefont
  {Matsumoto}}, \ and\ \bibinfo {author} {\bibfnamefont {Y.}~\bibnamefont
  {Omura}},\ }\bibfield  {title} {\enquote {\bibinfo {title}
  {Bernstein-greene-kruskal analysis of electrostatic solitary waves observed
  with geotail},}\ }\href {\doibase 10.1029/97JA02033} {\bibfield  {journal}
  {\bibinfo  {journal} {J. Geophys. Res. Space Phys.}\ }\textbf {\bibinfo
  {volume} {102}},\ \bibinfo {pages} {22131--22139} (\bibinfo {year}
  {1997})}\BibitemShut {NoStop}%
\bibitem [{\citenamefont {Omura}\ \emph {et~al.}(1996)\citenamefont {Omura},
  \citenamefont {Matsumoto}, \citenamefont {Miyake},\ and\ \citenamefont
  {Kojima}}]{omura1996electron}%
  \BibitemOpen
  \bibfield  {author} {\bibinfo {author} {\bibfnamefont {Y.}~\bibnamefont
  {Omura}}, \bibinfo {author} {\bibfnamefont {H.}~\bibnamefont {Matsumoto}},
  \bibinfo {author} {\bibfnamefont {T.}~\bibnamefont {Miyake}}, \ and\ \bibinfo
  {author} {\bibfnamefont {H.}~\bibnamefont {Kojima}},\ }\bibfield  {title}
  {\enquote {\bibinfo {title} {Electron beam instabilities as generation
  mechanism of electrostatic solitary waves in the magnetotail},}\ }\href
  {\doibase 10.1029/95JA03145} {\bibfield  {journal} {\bibinfo  {journal}
  {Journal of Geophysical Research: Space Physics}\ }\textbf {\bibinfo {volume}
  {101}},\ \bibinfo {pages} {2685--2697} (\bibinfo {year} {1996})}\BibitemShut
  {NoStop}%
\bibitem [{\citenamefont {Mottez}(2001)}]{mottez2001instabilities}%
  \BibitemOpen
  \bibfield  {author} {\bibinfo {author} {\bibfnamefont {F.}~\bibnamefont
  {Mottez}},\ }\bibfield  {title} {\enquote {\bibinfo {title} {Instabilities
  and formation of coherent structures},}\ }\href {\doibase
  10.1023/A:1012224820136} {\bibfield  {journal} {\bibinfo  {journal}
  {Astrophys. Space Sci}\ }\textbf {\bibinfo {volume} {277}},\ \bibinfo {pages}
  {59--70} (\bibinfo {year} {2001})}\BibitemShut {NoStop}%
\bibitem [{\citenamefont {Umeda}, \citenamefont {Omura},\ and\ \citenamefont
  {Matsumoto}(2004)}]{umeda2004two}%
  \BibitemOpen
  \bibfield  {author} {\bibinfo {author} {\bibfnamefont {T.}~\bibnamefont
  {Umeda}}, \bibinfo {author} {\bibfnamefont {Y.}~\bibnamefont {Omura}}, \ and\
  \bibinfo {author} {\bibfnamefont {H.}~\bibnamefont {Matsumoto}},\ }\bibfield
  {title} {\enquote {\bibinfo {title} {Two-dimensional particle simulation of
  electromagnetic field signature associated with electrostatic solitary
  waves},}\ }\href {\doibase 10.1029/2003JA010000} {\bibfield  {journal}
  {\bibinfo  {journal} {J. Geophys. Res. Space Phys.}\ }\textbf {\bibinfo
  {volume} {109}} (\bibinfo {year} {2004}),\ 10.1029/2003JA010000}\BibitemShut
  {NoStop}%
\bibitem [{\citenamefont {Lu}\ \emph {et~al.}(2008)\citenamefont {Lu},
  \citenamefont {Lembege}, \citenamefont {Tao},\ and\ \citenamefont
  {Wang}}]{lu2008perpendicular}%
  \BibitemOpen
  \bibfield  {author} {\bibinfo {author} {\bibfnamefont {Q.}~\bibnamefont
  {Lu}}, \bibinfo {author} {\bibfnamefont {B.}~\bibnamefont {Lembege}},
  \bibinfo {author} {\bibfnamefont {J.}~\bibnamefont {Tao}}, \ and\ \bibinfo
  {author} {\bibfnamefont {S.}~\bibnamefont {Wang}},\ }\bibfield  {title}
  {\enquote {\bibinfo {title} {Perpendicular electric field in two-dimensional
  electron phase-holes: A parameter study},}\ }\href@noop {} {\bibfield
  {journal} {\bibinfo  {journal} {Journal of Geophysical Research: Space
  Physics}\ }\textbf {\bibinfo {volume} {113}} (\bibinfo {year}
  {2008})}\BibitemShut {NoStop}%
\bibitem [{\citenamefont {Muschietti}\ \emph {et~al.}(2000)\citenamefont
  {Muschietti}, \citenamefont {Roth}, \citenamefont {Carlson},\ and\
  \citenamefont {Ergun}}]{muschietti2000transverse}%
  \BibitemOpen
  \bibfield  {author} {\bibinfo {author} {\bibfnamefont {L.}~\bibnamefont
  {Muschietti}}, \bibinfo {author} {\bibfnamefont {I.}~\bibnamefont {Roth}},
  \bibinfo {author} {\bibfnamefont {C.}~\bibnamefont {Carlson}}, \ and\
  \bibinfo {author} {\bibfnamefont {R.}~\bibnamefont {Ergun}},\ }\bibfield
  {title} {\enquote {\bibinfo {title} {Transverse instability of magnetized
  electron holes},}\ }\href@noop {} {\bibfield  {journal} {\bibinfo  {journal}
  {Physical Review Letters}\ }\textbf {\bibinfo {volume} {85}},\ \bibinfo
  {pages} {94} (\bibinfo {year} {2000})}\BibitemShut {NoStop}%
\bibitem [{\citenamefont {Muschietti}\ \emph {et~al.}(2002)\citenamefont
  {Muschietti}, \citenamefont {Roth}, \citenamefont {Carlson},\ and\
  \citenamefont {Berthomier}}]{muschietti2002modeling}%
  \BibitemOpen
  \bibfield  {author} {\bibinfo {author} {\bibfnamefont {L.}~\bibnamefont
  {Muschietti}}, \bibinfo {author} {\bibfnamefont {I.}~\bibnamefont {Roth}},
  \bibinfo {author} {\bibfnamefont {C.}~\bibnamefont {Carlson}}, \ and\
  \bibinfo {author} {\bibfnamefont {M.}~\bibnamefont {Berthomier}},\ }\bibfield
   {title} {\enquote {\bibinfo {title} {Modeling stretched solitary waves along
  magnetic field lines},}\ }\href {\doibase 10.5194/npg-9-101-2002, 2002}
  {\bibfield  {journal} {\bibinfo  {journal} {Nonlinear Process. Geophys}\
  }\textbf {\bibinfo {volume} {9}},\ \bibinfo {pages} {101--109} (\bibinfo
  {year} {2002})}\BibitemShut {NoStop}%
\bibitem [{\citenamefont {Chen}\ and\ \citenamefont
  {Parks}(2001)}]{chen2001trapped}%
  \BibitemOpen
  \bibfield  {author} {\bibinfo {author} {\bibfnamefont {L.-J.}\ \bibnamefont
  {Chen}}\ and\ \bibinfo {author} {\bibfnamefont {G.~K.}\ \bibnamefont
  {Parks}},\ }\bibfield  {title} {\enquote {\bibinfo {title} {Trapped and
  passing electrons in bgk solitary waves},}\ }\href@noop {} {\bibfield
  {journal} {\bibinfo  {journal} {arXiv preprint physics/0103020}\ } (\bibinfo
  {year} {2001})}\BibitemShut {NoStop}%
\bibitem [{\citenamefont {Turikov}(1984)}]{turikov1984electron}%
  \BibitemOpen
  \bibfield  {author} {\bibinfo {author} {\bibfnamefont {V.}~\bibnamefont
  {Turikov}},\ }\bibfield  {title} {\enquote {\bibinfo {title} {Electron phase
  space holes as localized bgk solutions},}\ }\href {\doibase
  10.1088/0031-8949/30/1/015} {\bibfield  {journal} {\bibinfo  {journal} {Phys.
  Scr.}\ }\textbf {\bibinfo {volume} {30}},\ \bibinfo {pages} {73} (\bibinfo
  {year} {1984})}\BibitemShut {NoStop}%
\bibitem [{\citenamefont {Muschietti}\ \emph {et~al.}(1999)\citenamefont
  {Muschietti}, \citenamefont {Ergun}, \citenamefont {Roth},\ and\
  \citenamefont {Carlson}}]{muschietti1999phase}%
  \BibitemOpen
  \bibfield  {author} {\bibinfo {author} {\bibfnamefont {L.}~\bibnamefont
  {Muschietti}}, \bibinfo {author} {\bibfnamefont {R.}~\bibnamefont {Ergun}},
  \bibinfo {author} {\bibfnamefont {I.}~\bibnamefont {Roth}}, \ and\ \bibinfo
  {author} {\bibfnamefont {C.}~\bibnamefont {Carlson}},\ }\bibfield  {title}
  {\enquote {\bibinfo {title} {Phase-space electron holes along magnetic field
  lines},}\ }\href {\doibase 10.1029/1999GL900207} {\bibfield  {journal}
  {\bibinfo  {journal} {Geophys. Res. Lett}\ }\textbf {\bibinfo {volume}
  {26}},\ \bibinfo {pages} {1093--1096} (\bibinfo {year} {1999})}\BibitemShut
  {NoStop}%
\bibitem [{\citenamefont {Goldman}, \citenamefont {Newman},\ and\ \citenamefont
  {Mangeney}(2007)}]{goldman2007theory}%
  \BibitemOpen
  \bibfield  {author} {\bibinfo {author} {\bibfnamefont {M.~V.}\ \bibnamefont
  {Goldman}}, \bibinfo {author} {\bibfnamefont {D.~L.}\ \bibnamefont {Newman}},
  \ and\ \bibinfo {author} {\bibfnamefont {A.}~\bibnamefont {Mangeney}},\
  }\bibfield  {title} {\enquote {\bibinfo {title} {Theory of weak bipolar
  fields and electron holes with applications to space plasmas},}\ }\href@noop
  {} {\bibfield  {journal} {\bibinfo  {journal} {Physical review letters}\
  }\textbf {\bibinfo {volume} {99}},\ \bibinfo {pages} {145002} (\bibinfo
  {year} {2007})}\BibitemShut {NoStop}%
\bibitem [{\citenamefont {Franz}\ \emph {et~al.}(2000)\citenamefont {Franz},
  \citenamefont {Kintner}, \citenamefont {Seyler}, \citenamefont {Pickett},\
  and\ \citenamefont {Scudder}}]{franz2000perpendicular}%
  \BibitemOpen
  \bibfield  {author} {\bibinfo {author} {\bibfnamefont {J.}~\bibnamefont
  {Franz}}, \bibinfo {author} {\bibfnamefont {P.}~\bibnamefont {Kintner}},
  \bibinfo {author} {\bibfnamefont {C.}~\bibnamefont {Seyler}}, \bibinfo
  {author} {\bibfnamefont {J.}~\bibnamefont {Pickett}}, \ and\ \bibinfo
  {author} {\bibfnamefont {J.}~\bibnamefont {Scudder}},\ }\bibfield  {title}
  {\enquote {\bibinfo {title} {On the perpendicular scale of electron
  phase-space holes},}\ }\href {\doibase 10.1029/1999GL010733} {\bibfield
  {journal} {\bibinfo  {journal} {Geophysical research letters}\ }\textbf
  {\bibinfo {volume} {27}},\ \bibinfo {pages} {169--172} (\bibinfo {year}
  {2000})}\BibitemShut {NoStop}%
\bibitem [{\citenamefont {Ng}, \citenamefont {Bhattacharjee},\ and\
  \citenamefont {Skiff}(2006)}]{ng2006weakly}%
  \BibitemOpen
  \bibfield  {author} {\bibinfo {author} {\bibfnamefont {C.}~\bibnamefont
  {Ng}}, \bibinfo {author} {\bibfnamefont {A.}~\bibnamefont {Bhattacharjee}}, \
  and\ \bibinfo {author} {\bibfnamefont {F.}~\bibnamefont {Skiff}},\ }\bibfield
   {title} {\enquote {\bibinfo {title} {Weakly collisional landau damping and
  three-dimensional bernstein-greene-kruskal modes: New results on old
  problems},}\ }\href@noop {} {\bibfield  {journal} {\bibinfo  {journal}
  {Physics of plasmas}\ }\textbf {\bibinfo {volume} {13}},\ \bibinfo {pages}
  {055903} (\bibinfo {year} {2006})}\BibitemShut {NoStop}%
\bibitem [{\citenamefont {Schamel}(1979)}]{schamel1979theory}%
  \BibitemOpen
  \bibfield  {author} {\bibinfo {author} {\bibfnamefont {H.}~\bibnamefont
  {Schamel}},\ }\bibfield  {title} {\enquote {\bibinfo {title} {Theory of
  electron holes},}\ }\href {\doibase 10.1088/0031-8949/20/3-4/006} {\bibfield
  {journal} {\bibinfo  {journal} {Phys. Scr.}\ }\textbf {\bibinfo {volume}
  {20}},\ \bibinfo {pages} {336} (\bibinfo {year} {1979})}\BibitemShut
  {NoStop}%
\bibitem [{\citenamefont {Chen}\ and\ \citenamefont
  {Parks}(2002)}]{chen2002GRL}%
  \BibitemOpen
  \bibfield  {author} {\bibinfo {author} {\bibfnamefont {L.-J.}\ \bibnamefont
  {Chen}}\ and\ \bibinfo {author} {\bibfnamefont {G.~K.}\ \bibnamefont
  {Parks}},\ }\bibfield  {title} {\enquote {\bibinfo {title} {Bgk electron
  solitary waves in 3d magnetized plasma},}\ }\href {\doibase
  https://doi.org/10.1029/2001GL013385} {\bibfield  {journal} {\bibinfo
  {journal} {Geophysical Research Letters}\ }\textbf {\bibinfo {volume} {29}},\
  \bibinfo {pages} {45--1--45--4} (\bibinfo {year} {2002})}\BibitemShut
  {NoStop}%
\bibitem [{\citenamefont {Hutchinson}(2021)}]{hutchinson2021oblate}%
  \BibitemOpen
  \bibfield  {author} {\bibinfo {author} {\bibfnamefont {I.}~\bibnamefont
  {Hutchinson}},\ }\bibfield  {title} {\enquote {\bibinfo {title} {Oblate
  electron holes are not attributable to anisotropic shielding},}\ }\href
  {\doibase 10.1063/5.0039233} {\bibfield  {journal} {\bibinfo  {journal}
  {Phys. Plasmas}\ }\textbf {\bibinfo {volume} {28}},\ \bibinfo {pages}
  {022902} (\bibinfo {year} {2021})}\BibitemShut {NoStop}%
\bibitem [{\citenamefont {Vasko}\ \emph {et~al.}(2017)\citenamefont {Vasko},
  \citenamefont {Agapitov}, \citenamefont {Mozer}, \citenamefont {Artemyev},
  \citenamefont {Drake},\ and\ \citenamefont {Kuzichev}}]{vasko2017electron}%
  \BibitemOpen
  \bibfield  {author} {\bibinfo {author} {\bibfnamefont {I.}~\bibnamefont
  {Vasko}}, \bibinfo {author} {\bibfnamefont {O.}~\bibnamefont {Agapitov}},
  \bibinfo {author} {\bibfnamefont {F.}~\bibnamefont {Mozer}}, \bibinfo
  {author} {\bibfnamefont {A.}~\bibnamefont {Artemyev}}, \bibinfo {author}
  {\bibfnamefont {J.}~\bibnamefont {Drake}}, \ and\ \bibinfo {author}
  {\bibfnamefont {I.}~\bibnamefont {Kuzichev}},\ }\bibfield  {title} {\enquote
  {\bibinfo {title} {Electron holes in the outer radiation belt:
  Characteristics and their role in electron energization},}\ }\href {\doibase
  10.1002/2016JA023083} {\bibfield  {journal} {\bibinfo  {journal} {J. Geophys.
  Res. Space Phys.}\ }\textbf {\bibinfo {volume} {122}},\ \bibinfo {pages}
  {120--135} (\bibinfo {year} {2017})}\BibitemShut {NoStop}%
\bibitem [{\citenamefont {Landau}\ and\ \citenamefont
  {Lifshitz}(1976)}]{landau1976mechanics}%
  \BibitemOpen
  \bibfield  {author} {\bibinfo {author} {\bibfnamefont {L.~D.}\ \bibnamefont
  {Landau}}\ and\ \bibinfo {author} {\bibfnamefont {E.~M.}\ \bibnamefont
  {Lifshitz}},\ }\href@noop {} {\emph {\bibinfo {title} {Mechanics}}},\
  Vol.~\bibinfo {volume} {1}\ (\bibinfo  {publisher} {Butterworth-Heinemann},\
  \bibinfo {year} {1976})\BibitemShut {NoStop}%
\bibitem [{\citenamefont {Goldman}\ \emph {et~al.}(2000)\citenamefont
  {Goldman}, \citenamefont {Crary}, \citenamefont {Newman},\ and\ \citenamefont
  {Oppenheim}}]{goldman2000turbulence}%
  \BibitemOpen
  \bibfield  {author} {\bibinfo {author} {\bibfnamefont {M.~V.}\ \bibnamefont
  {Goldman}}, \bibinfo {author} {\bibfnamefont {F.}~\bibnamefont {Crary}},
  \bibinfo {author} {\bibfnamefont {D.~L.}\ \bibnamefont {Newman}}, \ and\
  \bibinfo {author} {\bibfnamefont {M.}~\bibnamefont {Oppenheim}},\ }\bibfield
  {title} {\enquote {\bibinfo {title} {Turbulence driven by two-stream
  instability in a magnetized plasma},}\ }\href@noop {} {\bibfield  {journal}
  {\bibinfo  {journal} {Physics of Plasmas}\ }\textbf {\bibinfo {volume} {7}},\
  \bibinfo {pages} {1732--1739} (\bibinfo {year} {2000})}\BibitemShut {NoStop}%
\bibitem [{\citenamefont {Umeda}\ \emph {et~al.}(2006)\citenamefont {Umeda},
  \citenamefont {Omura}, \citenamefont {Miyake}, \citenamefont {Matsumoto},\
  and\ \citenamefont {Ashour-Abdalla}}]{umeda2006nonlinear}%
  \BibitemOpen
  \bibfield  {author} {\bibinfo {author} {\bibfnamefont {T.}~\bibnamefont
  {Umeda}}, \bibinfo {author} {\bibfnamefont {Y.}~\bibnamefont {Omura}},
  \bibinfo {author} {\bibfnamefont {T.}~\bibnamefont {Miyake}}, \bibinfo
  {author} {\bibfnamefont {H.}~\bibnamefont {Matsumoto}}, \ and\ \bibinfo
  {author} {\bibfnamefont {M.}~\bibnamefont {Ashour-Abdalla}},\ }\bibfield
  {title} {\enquote {\bibinfo {title} {Nonlinear evolution of the electron
  two-stream instability: Two-dimensional particle simulations},}\ }\href@noop
  {} {\bibfield  {journal} {\bibinfo  {journal} {Journal of Geophysical
  Research: Space Physics}\ }\textbf {\bibinfo {volume} {111}} (\bibinfo {year}
  {2006})}\BibitemShut {NoStop}%
\bibitem [{\citenamefont {Schamel}(1986)}]{schamel1986tutorial}%
  \BibitemOpen
  \bibfield  {author} {\bibinfo {author} {\bibfnamefont {H.}~\bibnamefont
  {Schamel}},\ }\bibfield  {title} {\enquote {\bibinfo {title} {Electron holes,
  ion holes and double layers: Electrostatic phase space structures in theory
  and experiment},}\ }\href {\doibase 10.1016/0370-1573(86)90043-8} {\bibfield
  {journal} {\bibinfo  {journal} {Physics Reports}\ }\textbf {\bibinfo {volume}
  {140}},\ \bibinfo {pages} {161--191} (\bibinfo {year} {1986})}\BibitemShut
  {NoStop}%
\bibitem [{\citenamefont {Hutchinson}(2017)}]{hutchinson2017electron}%
  \BibitemOpen
  \bibfield  {author} {\bibinfo {author} {\bibfnamefont {I.~H.}\ \bibnamefont
  {Hutchinson}},\ }\bibfield  {title} {\enquote {\bibinfo {title} {Electron
  holes in phase space: What they are and why they matter},}\ }\href {\doibase
  10.1063/1.4976854} {\bibfield  {journal} {\bibinfo  {journal} {Phys.
  Plasmas}\ }\textbf {\bibinfo {volume} {24}},\ \bibinfo {pages} {055601}
  (\bibinfo {year} {2017})}\BibitemShut {NoStop}%
\bibitem [{\citenamefont {Oppenheim}, \citenamefont {Newman},\ and\
  \citenamefont {Goldman}(1999)}]{oppenheim1999evolution}%
  \BibitemOpen
  \bibfield  {author} {\bibinfo {author} {\bibfnamefont {M.}~\bibnamefont
  {Oppenheim}}, \bibinfo {author} {\bibfnamefont {D.}~\bibnamefont {Newman}}, \
  and\ \bibinfo {author} {\bibfnamefont {M.}~\bibnamefont {Goldman}},\
  }\bibfield  {title} {\enquote {\bibinfo {title} {Evolution of electron
  phase-space holes in a 2d magnetized plasma},}\ }\href@noop {} {\bibfield
  {journal} {\bibinfo  {journal} {Physical review letters}\ }\textbf {\bibinfo
  {volume} {83}},\ \bibinfo {pages} {2344} (\bibinfo {year}
  {1999})}\BibitemShut {NoStop}%
\bibitem [{\citenamefont {Burch}\ \emph {et~al.}(2016)\citenamefont {Burch},
  \citenamefont {Torbert}, \citenamefont {Phan}, \citenamefont {Chen},
  \citenamefont {Moore}, \citenamefont {Ergun}, \citenamefont {Eastwood},
  \citenamefont {Gershman}, \citenamefont {Cassak}, \citenamefont {Argall}
  \emph {et~al.}}]{burch2016electron}%
  \BibitemOpen
  \bibfield  {author} {\bibinfo {author} {\bibfnamefont {J.}~\bibnamefont
  {Burch}}, \bibinfo {author} {\bibfnamefont {R.}~\bibnamefont {Torbert}},
  \bibinfo {author} {\bibfnamefont {T.}~\bibnamefont {Phan}}, \bibinfo {author}
  {\bibfnamefont {L.-J.}\ \bibnamefont {Chen}}, \bibinfo {author}
  {\bibfnamefont {T.}~\bibnamefont {Moore}}, \bibinfo {author} {\bibfnamefont
  {R.}~\bibnamefont {Ergun}}, \bibinfo {author} {\bibfnamefont
  {J.}~\bibnamefont {Eastwood}}, \bibinfo {author} {\bibfnamefont
  {D.}~\bibnamefont {Gershman}}, \bibinfo {author} {\bibfnamefont
  {P.}~\bibnamefont {Cassak}}, \bibinfo {author} {\bibfnamefont
  {M.}~\bibnamefont {Argall}},  \emph {et~al.},\ }\bibfield  {title} {\enquote
  {\bibinfo {title} {Electron-scale measurements of magnetic reconnection in
  space},}\ }\href {\doibase 10.1126/science.aaf2939} {\bibfield  {journal}
  {\bibinfo  {journal} {Science}\ }\textbf {\bibinfo {volume} {352}} (\bibinfo
  {year} {2016}),\ 10.1126/science.aaf2939}\BibitemShut {NoStop}%
\end{thebibliography}%
\providecommand{\noopsort}[1]{}\providecommand{\singleletter}[1]{#1}%

\end{document}